\documentclass[12pt,a4paper]{article}
\usepackage{epsfig}
\usepackage{mathtools}
\usepackage{amsfonts}
\usepackage{amssymb}
\usepackage{amsmath}
\usepackage{color}
\usepackage{slashed}
\usepackage{cite}
\usepackage{caption}
\usepackage{subcaption}

\usepackage{tikz}
\usetikzlibrary{shapes}
\usetikzlibrary{trees}
\usetikzlibrary{calc}
\usetikzlibrary{decorations.pathmorphing}
\usetikzlibrary{decorations.markings}

\def\la              {\langle}
\def\ra              {\rangle}

\newcommand{\dotalpha}{{\dot{\alpha}}}

\newcommand{\refspina}{\xi_A}
\newcommand{\refspinb}{\xi_B}
\newcommand{\refspinal}{\xi_A}
\newcommand{\refspinbl}{\xi_B}

\newcommand{\abr}[1]{\langle #1 \rangle}

\newcommand{\vll}{{\smash{\lambda}}}
\newcommand{\vlt}{{\smash{\tilde{\lambda}}}}
\newcommand{\vlet}{{\smash{\tilde{\eta}}}}

\newcommand{\vltu}{\smash{\underline{\smash{\tilde{\lambda}}}}}
\newcommand{\vleu}{\smash{\underline{\smash{\tilde{\eta}}}} }

\newcommand{\vlluu}{\smash{\underline{\underline{\smash{\lambda}}}}}
\newcommand{\vltuu}{\smash{\underline{\underline{\smash{\tilde{\lambda}}}}}}
\newcommand{\vleuu}{\smash{\underline{\underline{\smash{\tilde{\eta}}}}}}
\newcommand{\vletuu}{\smash{\underline{\underline{\smash{\eta}}}}}

\newcommand{\eqncom}{\, , }

\newcommand{\splus}{\! + \!}

\newcommand{\ssep}{\;}

\definecolor{grayn}{gray}{0.7}
\definecolor{lightgrayn}{gray}{0.8}

\def\vacuumheight{1}

\def\labelvdist{0.3}

\def\labelddist{\labelvdist*0.70710678118}

\newlength{\vacuumradius}
\newlength{\onshellradius}
\tikzstyle{db}=[circle, black, fill=black, minimum width=\onshellradius, draw, inner sep=0pt]
\tikzstyle{dw}=[circle, black, fill=white, minimum width=\onshellradius, draw, inner sep=0pt]
\tikzstyle{dvac}=[circle, black, fill=lightgrayn, minimum width=\vacuumradius, inner sep=0pt]
\tikzstyle{dl}=[circle, black, fill=white, inner sep=2pt]

\tikzset{
    gluon/.style={decorate, decoration={coil, amplitude=4pt, segment length=7pt, aspect=1}, draw=black}
}

\tikzset{->-/.style={decoration={
			markings,
			mark=at position .5 with {\arrow{>}}},postaction={decorate}}}

\newcommand{\C}{\mathbb{C}}
\newcommand{\T}{\mathbb{T}}

\newcommand{\CP}{\mathbb{CP}}
\newcommand{\PA}{\mathbb{PA}}
\newcommand{\PT}{\mathbb{PT}}

\newcommand{\bigslant}[2]{{\raisebox{.2em}{$#1$}\left/\raisebox{-.2em}{$#2$}\right.}}

\def\d{\text{d}}
\def\e{\text{e}}
\newcommand{\dbar}{\bar\partial}
\newcommand{\g}{\mathfrak{g}}

\begin{document}

\begin{center}

\vspace{1cm}

{\bf \Large Four dimensional ambitwistor strings  and \\ form factors of local and Wilson line operators.} \vspace{1cm}

{\large L.V. Bork$^{1,2,3}$ A.I. Onishchenko$^{4,5,6}$}\vspace{0.5cm}

{\it $^1$Institute for Theoretical and Experimental Physics, Moscow,
	Russia,\\
	$^2$National Research Nuclear University (MEPhI), Moscow,
	Russia,\\
	$^3$The Center for Fundamental and Applied Research, All-Russia
	Research Institute of Automatics, Moscow, Russia, \\
	$^4$Bogoliubov Laboratory of Theoretical Physics, Joint
	Institute for Nuclear Research, Dubna, Russia, \\
	$^5$Moscow Institute of Physics and Technology (State University), Dolgoprudny, Russia, \\
	$^6$Skobeltsyn Institute of Nuclear Physics, Moscow State University, Moscow, Russia}\vspace{1cm}

\abstract{We consider the description of form factors of local and Wilson line operators (reggeon amplitudes) in $\mathcal{N}=4$ SYM within the framework of four dimensional ambitwistor string theory. We present the explicit expressions for string composite operators corresponding to stress-tensor operator supermultiplet and Wilson line operator insertion. It is shown, that corresponding tree-level string correlation functions correctly reproduce previously obtained Grassmannian integral representations. As by product we derive four dimensional tree-level scattering equations representations for the mentioned form factors and formulate a simple gluing procedure used to relate operator form factors with on-shell amplitudes.
}
\end{center}

\begin{center}
Keywords: ambitwistor strings, super Yang-Mills theory, off-shell amplitudes, form factors, Wilson lines, superspace, reggeons
\end{center}

\newpage

\tableofcontents{}\vspace{0.5cm}

\renewcommand{\theequation}{\thesection.\arabic{equation}}

\section{Introduction}

Recently twistor string theories \cite{WittenTwistorStringTheory,BerkovitsTwistorStringTheory} played a crucial role in understanding and discovery of mathematical structures underlying scattering amplitudes in $\mathcal{N}=4$ super Yang-Mills and $\mathcal{N}=8$ supergravity in four dimensions.

Based on Witten's twistor string theory Roiban, Spradlin and Volovich (RSV) got the integral representation of  $\mathcal{N}=4$ SYM tree level N$^{k-2}$MHV amplitudes as integrals over the moduli space of degree $k-1$ curves in super twistor space  \cite{RSV,SpradlinVolovichFromTwistorString}. Further generalization of RSV result was performed by Cachazo, He and Yuan (CHY)  via the introduction of so called {\it scattering equations} \cite{scatteringEq1,scatteringEq2,scatteringEq3,scatteringEq4,scatteringEq5}.  Within the latter  $\mathcal{N}=4$ SYM amplitudes are expressed in terms of integrals over the marked points on the Riemann sphere, which are localized on the solutions of mentioned scattering equations. Next the CHY formulae where shown to come naturally from {\it ambitwistor string theory} \cite{AmbitwistorStringsScatEqns,ambitwistorString4d}, which was also used to obtain loop-level generalization of scattering equations representation \cite{AmbitwistorStringsScatEqns-1loop,WorldsheetGeometriesAmbitwistorString,InraredBehaviorScatEqns-1loop,ScatteringEqnsPureSpinors,LoopIntegrandsRiemannSphere,OneLoopRiemannSphere,EllipticScatEqns,CHYGraphsTorus,OneLoopFromTree,TwoLoopRiemannSphere}.
Another close direction in the study of scattering amplitudes is related to their representation in terms of integrals over Grassmannians
\cite{DualitySMatrix,AmplitudesPositiveGrassmannian,AllLoopIntegrandN4SYM,GrassmanianOriginDualConformalInvariance,UnificationResidues,DualSupercondormalInvarianceMomentumTwistorsGrassmannians}.

First, this representation allows natural unification of different BCFW \cite{BCFW1,BCFW2} representations for tree level amplitudes and loop level integrands  \cite{DualitySMatrix,AmplitudesPositiveGrassmannian}. Second, it is ultimately related to the integrable structure behind $\mathcal{N}=4$ SYM S-matrix \cite{DrummondYangianSymmetry,DrummondSuperconformalSymmetry,YangBaxterScatteringAmplitudes,BetheAnsatzYangianInvariants,BeisertYangianSymmetry}. Moreover, the Grassmannian integral representation also naturally relates perturbative $\mathcal{N}=4$ SYM and twistor string theories amplitudes\cite{UnificationResidues}. Finally, the Grassmannian integral representation of scattering amplitudes has led to the discovery of geometrical structure of $\mathcal{N}=4$ SYM S-matrix (so called Amplituhedron)
\cite{MomentumTwistors,Arcani_Hamed_Polytopes,Amplituhdron_1,Amplituhdron_2,Amplituhdron_3,Amplituhdron_4,Amplituhdron_5,Amplituhdron_6}.

All mentioned above results are extremely relevant not only from pure theoretical, but also from  more practical point of view. For example, these results provide us with relatively compact analytical expressions for $n$-point tree level amplitudes in gauge theories with $\mathcal{N} \leq 4$ SUSY (including QCD), which in turn could be used to compute corresponding loop level amplitudes (see for a review \cite{Henrietta_Amplitudes}). It is important to note that all mentioned above results was almost impossible to obtain by standard textbook Feynman diagram methods.

Described above ideas and methods can be applied not only to the scattering amplitudes of
on-shell sates (S-matrix elements) but to form factors of gauge invariant operators (local or none local) as well.
The form factors of local gauge invariant operators is quite developed topic in a literature, see \cite{GeneralizedUnitarityFormFactors,PolytopesFormFactors,FormFactorsPeriodicWilsonLoops,FormFactorsColorKinematic5loop,
FormFactorsYsystem1,FormFactorsYsystem2,DilatationOperatorFormFactors1,DilatationOperatorFormFactors2,DilatationOperatorFormFactors3} and references therein.
The general practice when studying form factors is to consider the case of local gauge invariant color singlet operators. However, one may also consider gauge invariant (the representation under global gauge transformation is not necessary singlet) non-local operators, for example  Wilson loops (lines) or their products \cite{vanHamerenBCFW1,vanHamerenBCFW2,vanHamerenBCFW3,LipatovEL1,LipatovEL2,KirschnerEL1,KirschnerEL2,
KotkoWilsonLines,vanHamerenWL1,vanHamerenWL2,offshell-1leg,offshell-multiplelegs,WilsonLoopFormFactors}.
An insertion of Wilson line operator will then correspond to the off-shell or reggeized gluon in such formulation. These objects should be more familiar to the reader as gauge invariant off-shell amplitudes  \cite{vanHamerenBCFW1,vanHamerenBCFW2,vanHamerenBCFW3,LipatovEL1,LipatovEL2,KirschnerEL1,KirschnerEL2,KotkoWilsonLines,vanHamerenWL1,vanHamerenWL2} (also known as reggeon amplitudes in the framework of Lipatov's effective lagrangian), which appear within the context of
$k_T$  or high-energy factorization
\cite{GribovLevinRyskin,CataniCiafaloniHautmann,CollinsEllis,CataniHautmann} as well as in the study of processes at multi-regge kinematics.

Up to the moment we already have scattering equations (connected prescription) representations for the form factors of operators from stress-tensor operator supermultiplet and  scalar operators of the form $\text{Tr}(\phi^m)$ \cite{BrandhuberConnectedPrescription,HeConnectedFormulaFormFactors}. Also the connected prescription formulae were extended to Standard Model amplitudes \cite{HeConnectedFormulaSM}. Besides, there are several results for the Grassmannian integral representation of form factors of operators from stress-tensor operator supermultiplet \cite{SoftTheoremsFormFactors,FormFactorsGrassmanians,WilhelmThesis,q2zeroFormFactors} and Wilson line operator insertions \cite{offshell-1leg,offshell-multiplelegs}, see also \cite{WilsonLoopFormFactors} for a recent interesting discussion of duality
for Wilson loop form factors
\footnote{A very close subject is the twistor and Lorentz harmonic chiral superspace formulation of form factors and correlation functions developed in \cite{TwistorFormFactors1,TwistorFormFactors2,TwistorFormFactors3,LHC1,LHC2,LHC3,LHC4}}.

The purpose of this work is further pursue the string based approach to $\mathcal{N}=4$ super Yang-Mills and other four dimensional gauge theories. Namely we want to derive Grassmannian integral and scattering equations representation \cite{FormFactorsGrassmanians,WilhelmThesis} and \cite{offshell-1leg,offshell-multiplelegs} of form factors and correlation functions of local and Wilson line operators in $\mathcal{N}=4$ SYM starting from four dimensional ambitwistor string theory. Recently we have already provided such a derivation for the case of reggeon amplitude (Wilson line form factors) in \cite{BorkOnishchenkoAmbitwistorReggeon}. This paper contains both extra details of the latter derivation together with its extension to the case of the form factors of local operators. In addition we further investigate relation between on-shell amplitudes
and form factors and suggest procedure which allows one to reconstruct (at tree and possible
loop level) form factors and correlation functions of
Wilson line operators starting directly from on-shell scattering amplitudes.

This paper is organized as follows: in section \ref{formfactorsSec} we introduce necessary definitions for the form factors of operators from stress-tensor supermultiplet and Wilson line operators.

In section \ref{4dAmbitwistorStringsSec}, to make article self contained, we proceed with the recalling of general four dimensional ambitwistor string theory formalism \cite{ambitwistorString4d}.

In section \ref{StringVertexesCorrelationFunctionsSec} we discuss our motivation
to introduce so called \emph{gluing procedure} -- an operation, introduced for the first time in \cite{BorkOnishchenkoAmbitwistorReggeon}, which is given by convolution of the ambitwistor
vertex operators with some function of external kinematical data. Based on this gluing procedure we present expressions for string theory generalized vertex operators (string theory composite operators) corresponding to $\mathcal{N}=4$ SYM field theory stress-tensor operator as well as we give some additional details regarding derivation of the results of \cite{BorkOnishchenkoAmbitwistorReggeon}. Using these new string vertex operators we compute corresponding tree-level string theory correlation functions and show, that they correctly reproduce the results of previously obtained Grassmannian integral representations of stress tensor supermultiplet form factors \cite{FormFactorsGrassmanians} as well as form factors and correlation functions of Wilson line operators \cite{offshell-1leg,offshell-multiplelegs}.

Section \ref{GrassmanniansLinkRepresentationSec} contains
detailed review based on \cite{UnificationResidues} about relation between Grassmannian integral and RSV (scattering equation) representations of the on-shell amplitudes, which plays very important role in our construction as well. In the end of the section we briefly discuss one simple self consistency check of our construction.

In section \ref{GluingAmplitudesSec} we further discuss gluing procedure. We show that
one can formally apply it (a convolution with some function of external kinematical data) directly on the level of on-shell amplitudes represented as
the sum of BCFW terms (both at tree and, probably, integrand level). We formalise this
by introducing notion of \emph{gluing operator} $\hat{A}$. Using gluing operator we reproduce several previously obtained \cite{vanHamerenBCFW1,offshell-1leg,offshell-multiplelegs} results for Wlison line form factors (reggeon amplitudes) including 3-point correlation function of reggeized gluons.

Finally, in section \ref{ConclusionSec} we present our conclusion and discuss possible future research directions. Appendices \ref{appA} and \ref{appC} contain the computational details of form factor gluing procedure.

\section{Form factors of local and  Wilson line operators}\label{formfactorsSec}
In this work we will be interested in ambitwistor string description of form factors of Wilson line and local operators in $\mathcal{N}=4$ SYM. $\mathcal{N}=4$ SYM is a maximally supersymmetric gauge theory in four space time dimensions \cite{N4SYM_1,N4SYM_2}. Its sixteen on-shell states (their creation/annihilation operators) could be conveniently described using $\mathcal{N}=4$ on-shell chiral superfield \cite{Nair}:
\begin{eqnarray}
\Omega= g^+ + \vlet_A\psi^A + \frac{1}{2!}\vlet_{A}\vlet_{B}\phi^{AB}
+ \frac{1}{3!}\vlet_A\vlet_B\vlet_C\epsilon^{ABCD}\bar{\psi}_{D}
+ \frac{1}{4!}\vlet_A\vlet_B\vlet_C\vlet_D\epsilon^{ABCD}g^{-},
\end{eqnarray}
Here, $g^+, g^-$ denote creation/annihilation operators of gluons with $+1$ and $-1$ hecilities, $\psi^A$ are creation/annihilation operators of four Weyl spinors with negative helicity $-1/2$, $\bar{\psi}_A$ are creation/annihilation operators of four Weyl spinors with positive helicity and $\phi^{AB}$ stand for creation/annihilation operators of six scalars (anti-symmetric in the $SU(4)_R$ $R$-symmetry group indices $AB$). All $\mathcal{N}=4$ SYM fields transform  in the adjoint representation of $SU(N_c)$ gauge group. In what follows we will also need superstates defined by the action of superfield creation/annihilation operators on vacuum. For $n$-particle superstate we have
\begin{eqnarray}
|\Omega_1\Omega_2\ldots\Omega_n\ra\equiv \Omega_1\Omega_2\ldots\Omega_n|0\ra
\end{eqnarray}

Form factors of Wilson line operators are generally used to describe gauge invariant off-shell or reggeon amplitudes \cite{vanHamerenBCFW1,vanHamerenBCFW2,vanHamerenBCFW3,LipatovEL1,LipatovEL2,KirschnerEL1,KirschnerEL2,KotkoWilsonLines,vanHamerenWL1,vanHamerenWL2}. The Wilson line operators used to describe off-shell reggeized gluons are defined as \cite{KotkoWilsonLines}:
\begin{eqnarray}\label{WilsonLineOperDef}
	\mathcal{W}_p^c(k) = \int d^4 x e^{ix\cdot k} \mathrm{Tr} \left\{
	\frac{1}{\pi g} t^c \; \mathcal{P} \exp\left[\frac{ig}{\sqrt{2}}\int_{-\infty}^{\infty}
	ds \; p\cdot A_b (x+ sp) t^b\right]
	\right\}.
\end{eqnarray}
where $t^c$ is $SU(N_c)$ generator\footnote{The color generators are normalized as $\mathrm{Tr} (t^a t^b) = \delta^{a b}$} and we also used so called  $k_T$ - decomposition of the off-shell gluon momentum $k$, $k^2 \neq 0$:
\begin{eqnarray}\label{kT}
	k^{\mu} = x p^{\mu} + k_T^{\mu} .
\end{eqnarray}
Here, $p$ is the off-shell gluon direction (also known as gluon polarization vector), such that $p^2=0$, $p\cdot k = 0$ and $x \in [0,1]$.  Such decomposition is generally parametrized by an auxiliary light-cone four-vector $q^{\mu}$, so that
\begin{eqnarray}
	k_T^{\mu} (q) = k^{\mu} - x(q) p^{\mu}\quad \text{with}\quad x(q) = \frac{q\cdot k}{q\cdot p} \;\; \text{and} \;\; q^2 = 0.
\end{eqnarray}
As momentum $k_T^{\mu}$ is transverse with respect to both $p^\mu$ and $q^\mu$ vectors one can decompose it into the basis of two ``polarization'' vectors\footnote{Here we used the helicity spinor decomposition of light-like four-vectors $p$ and $q$.}  as \cite{vanHamerenBCFW1}:
\begin{eqnarray}
	k_T^{\mu} (q) = -\frac{\kappa}{2}\frac{\la p|\gamma^{\mu}|q]}{[pq]}
	- \frac{\kappa^{*}}{2}\frac{\la q|\gamma^{\mu}|p]}{\la qp\ra}\quad
	\text{with} \quad \kappa = \frac{\la q|\slashed{k}|p]}{\la qp\ra},\;
	\kappa^{*} = \frac{\la p|\slashed{k}|q]}{[pq]}.
\end{eqnarray}
It is easy to see, that $k^2 = -\kappa\kappa^{*}$ and both $\kappa$ and $\kappa^{*}$ are independent of auxiliary four-vector $q^{\mu}$
\cite{vanHamerenBCFW1}. Another useful relation, which  is direct consequence of $k_T$ decomposition and will be used often in practical calculations later on, is
\begin{eqnarray}\label{pBracetsRelation}
	k|p\rangle=|p]\kappa^*.
\end{eqnarray}
Note, that Wilson line operator we use to describe off-shell gluon is colored. It is invariant $\delta\mathcal{W}_p^c(k) = 0$ under local infinitesimal gauge transformations $\delta A_{\mu} = [D_{\mu},\chi]$ with $\chi$ decreasing at $x\to\infty$.  At the same time it transforms under global adjoint transformations of $SU (N_c)$ with constant $\chi$ as \cite{LipatovEL1,LipatovEL2}:
\begin{eqnarray}
	\delta\mathcal{W}_p(k) = g [\mathcal{W}_p(k), \chi] .
\end{eqnarray}

The form factor of Wilson line operator or gauge invariant  amplitude with one off-shell and $n$ on-shell gluons is then given by \cite{KotkoWilsonLines}:
\begin{eqnarray}\label{AmplitudeOnelOffShellGluon}
	\mathcal{A}_{n+1} \left(1^{\pm},\ldots ,n^{\pm},g_{n+1}^*\right) = \la \{k_i, \epsilon_i, c_i\}_{i=1}^n|\mathcal{W}_p^{c_{n+1}}(k)|0\ra\, ,
\end{eqnarray}
where asterisk denotes off-shell gluon, while $p$, $k$, $c$ stand for its direction, momentum and color index.  Next $\la \{k_i, \epsilon_i, c_i\}_{i=1}^m|=\bigotimes_{i=1}^m\la k_i,\varepsilon_i, c_i|$ and $\la k_i,\varepsilon_i, c_i|$ denotes on-shell gluon state  with momentum $k_i$, polarization vector $\varepsilon_i^-$ or $\varepsilon_i^+$ and color index $c_i$. Also in the case when there is no confusion in the position of Wilson line operator insertion the latter will be labeled just by $g^*$. Form factors with multiple Wilson line insertions or amplitudes with multiple off-shell gluons can be represented in a similar fashion:
\begin{eqnarray}\label{AmplitudeSeveralOffShellGluons}
	\mathcal{A}_{m+n} \left(1^{\pm},\ldots ,m^{\pm},g_{m+1}^*,\ldots ,g_{n+m}^*\right) =
	\la \{k_i, \epsilon_i, c_i\}_{i=1}^m|\prod_{j=1}^n \mathcal{W}_{p_{j+m}}^{c_{j+m}}(k_{j+m})|0\ra,
\end{eqnarray}
where $p_{i+m}$ is the direction of the $i$'th ($i=1,...,n$) off-shell gluon and $k_{i+m}$ is its off-shell momentum.  As a function of kinematical variables this amplitude is written as
\begin{eqnarray}
\mathcal{A}_{m+n} \left(1^{\pm},\ldots ,g_{n+m}^*\right) =\mathcal{A}_{m+n}\left(\{\lambda_i,\tilde{\lambda}_i,\pm,c_i\}_{i=1}^{m};
\{k_j,\lambda_{p,j},\tilde{\lambda}_{p,j},c_j\}_{j=m+1}^{m+n}\right),
\end{eqnarray}
where $\lambda_{p,j},\tilde{\lambda}_{p,j}$ are spinors coming from helicity spinor decomposition of polarization vector of $j$'th reggeized gluon. In the case when only off-shell gluons are present (correlation function of Wilson line operator insertions) we have:
\begin{eqnarray}\label{CorrFunctionOffShellGluons}
	\mathcal{A}_{0+n} \left(g_1^*\ldots g_n^*\right) =
	\la 0|\mathcal{W}_{p_{1}}^{c_{1}}(k_{1})\ldots \mathcal{W}_{p_{n}}^{c_{n}}(k_{n})|0\ra.
\end{eqnarray}
Of course it is also possible to consider color ordered versions of Wilson line form factors, while  the original off-shell amplitudes (Wilson line form factors) are then recovered using color decomposition\footnote{See for example \cite{offshell-1leg,DixonReview}.}:
\begin{eqnarray}\label{ColourOrderedAmplitudeDefinition}
	\mathcal{A}_{n+m}^*(1^{\pm},\ldots ,m^{\pm},g_{m+1}^*,\ldots ,g_{n+m}^*) &=& g^{n-2}\sum_{\sigma \in S_{n+m}/Z_{n+m}} \text{tr}
	~(t^{a_{\sigma (1)}}\cdots t^{a_{\sigma (n+m)}})\times\nonumber\\
 &\times&A_{n+m}^* \left(\sigma(1^{\pm}),\ldots, \sigma(g_{n+m}^*)\right).
\end{eqnarray}
Note, that in the planar limit this decomposition is valid both for arbitrary tree and loop level amplitudes.

In the case of $\mathcal{N}=4$ SYM one may also consider other then gluons on-shell states from $\mathcal{N}=4$ supermultiplet.  The corresponding  $\mathcal{N}=4$ SYM superamplitudes are then given by
\begin{eqnarray}\label{AmplitudeSeveralOffShellGluonsSUSY}
A_{m+n}^* \left(\Omega_1,\ldots,\Omega_m,g_{m+1}^*,\ldots ,g_{n+m}^*\right) =
\la \Omega_1\ldots\Omega_m|
\prod_{j=1}^n\mathcal{W}_{p_{m+j}}(k_{m+j})|0\ra,
\end{eqnarray}
and the explicit dependence of $A_{m+n}^* \left(\Omega_1,\ldots,g_{m+n}^*\right)$ amplitude on kinematical variables takes the form
\begin{eqnarray}\label{AmplitudeSeveralOffShellGluonsArgumentsSUSY}
A_{m+n}^* \left(\Omega_1,\ldots,g_{m+n}^*\right) =A_{m+n}^*\left(\{\lambda_i,\tilde{\lambda}_i,\tilde{\eta}_i\}_{i=1}^{m};
\{k_j,\lambda_{p,j},\tilde{\lambda}_{p,j}\}_{j=m+1}^{m+n}\right).
\end{eqnarray}
The above superamplitude contains not only component amplitudes with on-shell gluons, but also all amplitudes with other on-shell states from $\mathcal{N}=4$ supermultiplet. The helicity spinors $\lambda_i,\tilde{\lambda}_i$ encode kinematics of on-shell states, while $\tilde{\eta}_i$ encodes their helicity content. Off-shell momentum $k_i$ and light-cone direction vector $p_i=\lambda_{p,i}\tilde{\lambda}_{p,i}$ encode information related to Wilson line operator insertion. So, in what follows we will be considering partially supersymmetrized version of amplitudes (\ref{AmplitudeSeveralOffShellGluons}) with on-shell states treated in supersymmetric manner, while Wilson line operators ("off-shell states") left unsupersymmetrized. The component amplitudes containing  gluons, scalars and fermions may then be extracted as coefficients in $\tilde{\eta}$ expansion of $A_{m+n}^*$ superamplitude similar to the case of ordinary on-shell amplitudes and super form factors.

While our present consideration should be applicable\footnote{See corresponding discussion in Conclusion.} not only to Wilson line but to arbitrary local operators, here for concreteness we will restrict ourselves to the case of operators from stress-tensor operator supermultiplet. When considering the latter the general practice is to focus on the chiral part of this multiplet. Using harmonic superspace approach \cite{N=4_Harmonic_SS,SuperCor1} it is given by \cite{HarmonyofFF_Brandhuber,BKV_SuperForm,SuperCor1,SuperFormFactorsHalfBPS}:
\begin{equation}
{\mathcal T} (x, \theta^+) = \text{tr} (\phi^{++}\phi^{++}) + \cdots + \frac{1}{3}(\theta^+)^4{\mathcal{L}} ,
\end{equation}
where $u_A^{+a}$, $u_A^{-a'}$ is a set of harmonic coordinates parameterizing coset $\frac{SU(4)}{SU(2) \times SU(2)' \times U(1)}$ and $\theta_{\alpha}^{+a} = \theta_{\alpha}^A u_A^{+a}$, $\theta_{\alpha}^{-a'} = \theta_{\alpha}^A u_A^{-a'}$. Here, $A$ is $SU(4)_R$ index, $a$ and $a'$ are $SU(2)$ indices and $\pm$ denote $U(1)$ charges. For example $\epsilon^{ab}\phi^{++}=\phi^{AB}u_A^{+a}u_A^{+b}$, where $\phi^{AB}$ is the scalar field from $\mathcal{N}=4$ lagrangian. The color ordered form factors
of operators from the chiral truncation of stress-tensor operator supermultiplet $F_n$ are then given by
\begin{eqnarray}\label{ChiralTrFF}
F_n(\Omega_1,\ldots,\Omega_n;\mathcal{T}) \equiv \langle\Omega_1\ldots\Omega_n|\mathcal{T}(q,\gamma^{-})|0\rangle
=F_n\left(\{\lambda_i,\tilde{\lambda}_i,\tilde{\eta}_i\}_{i=1}^n;\{q,\gamma^{-}\}\right) \, ,
\end{eqnarray}
where $\{\lambda_i,\tilde{\lambda}_i,\tilde{\eta}_i\}_{i=1}^n$ are kinematical and helicity data of the on-shell states, $q$ is the operator momentum and $\gamma^{-}$ parametrizes the operator content of the chiral part of $\mathcal{N}=4$ SYM stress-tensor operator supermultiplet. Here, we have also performed the Fourier  transformation from variables  $x,\theta^{+}$ to $q,\gamma^{-}$ \cite{HarmonyofFF_Brandhuber,BKV_SuperForm}.
The full physical form factor may then be restored from its color ordered version using standard color decomposition formula
\begin{eqnarray}\label{ColourOrderedFormFactorDefinition} \mathcal{F}_n(\Omega_1,\ldots,\Omega_n;\mathcal{T}) &=& g^{n-2}\sum_{\sigma \in S_n/Z_n} \text{tr}
	~(t^{a_{\sigma (1)}}\cdots t^{a_{\sigma (n)}})\times\nonumber\\
&\times&F_n (\sigma(\Omega_1),\ldots ,\sigma (\Omega_n);\mathcal{T}),
\end{eqnarray}
where $S_n/Z_n$ denotes all none cyclic permutations of $n$ objects.
As in the case of off-shell amplitudes this formula is valid both for arbitrary
tree and loop level form factors in the planar limit\footnote{At loop level one should take into account appropriate powers of t'Hooft coupling constant $g^2N_c$, which were suppressed here.}. At least at tree level
the form factors of full stress-tensor operator supermultiplet can be reconstructed if the explicit form of (\ref{ChiralTrFF}) is known \cite{HarmonyofFF_Brandhuber}.

\section{Four dimensional ambitwistor strings}
\label{4dAmbitwistorStringsSec}
\subsection{General formalism}

As was already mentioned in Introduction to describe form factors of local and Wilson line operators we will be using the four dimensional ambitwistor string theory originally formulated in \cite{ambitwistorString4d}.  Here for completeness we will discuss essential
details of dimensional ambitwistor string theory. Our presentation of this theory here closely follows \cite{ambitwistorString4d} and we refer the interested reader to this original paper and \cite{GeyerThesis} for further details.

The target space of four dimensional ambitwistor string is given by projective ambitwistor space $\PA$. The latter is the supersymmetrized space of complex null geodesics in a complexified Minkowski given by a quardric $Z\cdot W = 0$ inside the product of twistor and dual twistor spaces $\PT\times \PT^*$ quotient by relative scaling $Z\cdot\partial_Z-W\cdot \partial_W$:
\begin{equation}
\PA=\bigslant{\big\{(Z,W)\in \T\times \T^* |\, Z\cdot W=0\big\}}{\{Z\cdot\partial_Z-W\cdot \partial_W\}} \,.
\end{equation}
In the case of $\mathcal{N}$ supersymmetries $Z = (\vll_{\alpha}, \mu^{\dotalpha}, \chi^r)\in\T = \C^{4|\mathcal{N}}$,  $W=(\tilde \mu ,\tilde \lambda, \tilde \chi)\in\T^*$ and $Z\cdot W =\lambda_\alpha\tilde\mu^\alpha+\mu^{\dot\alpha}\tilde\lambda_{\dot\alpha}+\chi^r\tilde\chi_r$, where $\chi, \tilde{\chi}$ are fermionic, $\alpha=0,1$, $\dot\alpha=\dot 0,\dot 1$ and $r = 1,\ldots ,\mathcal{N}$ is $\text{R}$-symmetry index. The point $(x, \theta ,\tilde{\theta})$ in non-chiral super Minkowski space corresponds to a quardric $\CP^1\times\CP^1$ parametrized by $(\vll,\vlt)$ spinors. The correspondence is realized by the standard twistor incidence relations
\begin{align}
&\mu^{\dot \alpha}=i(x^{\alpha\dot\alpha} +i\theta^{r\alpha}\tilde\theta^{\dot\alpha}_r)\lambda_\alpha\, ,&&  \chi^r=\theta^{r\alpha}\lambda_\alpha\, ,\\
&\tilde \mu^{ \alpha}=-i(x^{\alpha\dot\alpha} -i\theta^{r\alpha}\tilde\theta^{\dot\alpha}_r)\tilde \lambda_{\dot \alpha}\, ,&&
\tilde \chi_r=\tilde \theta_r^{\dot \alpha}\tilde \lambda_{\dot \alpha}\, ,
\end{align}
It is easy to check, that this quardric lies in $Z\cdot W = 0$

The four dimensional ambitwistor string consists from worldsheet spinors $(Z, W)$ with values in $\T\times\T^*$ and $\text{GL}(1, \C)$ gauge field $a$ acting as a Lagrange multiplier for the constraint $Z\cdot W = 0$. In the conformal gauge the action is given by\footnote{It is obtained by chiral pullback of contact stucture on ambitwistor space $\Theta = \frac i2(Z\cdot \d W-W\cdot \d Z)$ \cite{ambitwistorString4d}. Note, that similar action first appeared in\cite{BerkovitsTwistorStringTheory} in the context of open twistor string theory.}
\begin{equation}
S=\frac1{2\pi}\int_\Sigma  W\cdot \dbar Z-Z\cdot \dbar W  +a Z\cdot W +S_J\, , \label{stringAction}
\end{equation}
where $\dbar = \d\bar{\sigma}\partial_{\bar{\sigma}}$ ($\sigma, \bar{\sigma}$ are some local holomorphic and anti-holomorphic coordinates on Riemann surface $\Sigma$) and $S_J$ is the action for a worldsheet Kac-Moody current algebra  $J\in\Omega^0(\Sigma, K_\Sigma\otimes\g)$ for some Lie algebra $\g$. Here $K_{\Sigma}$ denotes canonical bundle on surface $\Sigma$ and the remaining worldsheet fields take values in
\begin{align}
&Z\in\Omega^0(\Sigma, K_\Sigma^{1/2}\otimes\T)\,,\\
&W\in\Omega^0(\Sigma,K_\Sigma^{1/2}\otimes\T^*)\,,\\
&a\in\Omega^{0,1}(\Sigma)\,,
\end{align}
where powers of canonical bundle denote fields conformal weights. The above action is invariant under a gauge symmetry
\begin{equation}
Z^I\rightarrow e^\gamma Z^I\,,\quad W_I\rightarrow e^{-\gamma}W_I\,,\quad a\rightarrow a-2\dbar \gamma\,,
\end{equation}
that quotients the target space by the action of $Z\cdot\partial_Z-W\cdot \partial_W$. The gauge fixing of worldsheet diffeomorphism symmetry\footnote{In a general gauge, the $\dbar$ operator in (\ref{stringAction}) is replaced by operator $\dbar_{\tilde{e}} = \dbar + \tilde{e}\partial$ parametrizing the worldsheet diffeomorphism freedom.} and the above gauge redundancy via standard BRST procedure leads to the introduction of the standard reparametrization (Virasoro) $(b,c)$  together with $\text{GL}(1)$ $(u, v)$ ghost systems:
\begin{align}
&c\in\Pi\Omega^0(\Sigma,T_\Sigma)\,, && v\in \Pi\Omega^0(\Sigma)\,,\\
&b\in\Pi\Omega^0(\Sigma,K_\Sigma^{2})\,, && u\in \Pi\Omega^0(\Sigma,K_\Sigma)\,,
\end{align}
where $T_\Sigma$ denotes tangent bundle on surface $\Sigma$ and $\Pi\Omega^0(\Sigma , E)$ denotes the space of fermion-valued sections of $E$. The full worldsheet action is then given by
\begin{equation}
S=\frac1{2\pi}\int_\Sigma  W\cdot \dbar Z-Z\cdot \dbar W  +b\dbar c+u\dbar v +S_J\,,
\end{equation}
and the BRST operator takes the form
\begin{equation}
Q=\oint c T + vZ\cdot W+Q_{\text{gh}}\,.
\end{equation}
where $T=W\cdot \partial Z-Z\cdot \partial W +T_J$ is the world-sheet stress-energy tensor.

\subsection{String vertex operators and their correlation functions}

To calculate string scattering amplitudes we need vertex operators. In general they are given by first-quantized wave functions of external states translated into worldsheet operator insertions. Penrose transform allows us to relate solutions to massless field equations in Minkowski space to cohomology classes on projective twistor space. In the case of Yang-Mills theory ambitwistor string vertex operators are obtained by pairing pullbacks of general ambitwistor space wave functions $\alpha\in H^1 (\PA , \mathcal{O})$ ($\dbar$-closed worldsheet $(0,1)$-forms) with Kac-Moody currents $J\cdot T_a$ to get $\mathcal{V}_a = \int_{\Sigma}\alpha\, J\cdot T_a$. For two types of momentum eigenstates (pullbacks from either twistor or dual twistor space) we get \cite{ambitwistorString4d}:
\begin{align}\label{VertexOperatorsOnShellStates}
&\mathcal{V}_a=\int\frac{\d s_{a}}{s_{a}}\bar{\delta}^{2}(\lambda_{a}-s_{a}\lambda)\e^{is_{a}\left([\mu\,\tilde{\lambda}_{a}]+\chi^{r}\tilde{\eta}_{ar}\right)} J\cdot T_a \,, \\
&\widetilde{\mathcal{V'}}_a= \int\frac{\d s_{a}}{s_{a}}\bar{\delta}^{2}(\tilde{\lambda}_{a}-s_{a}\tilde{\lambda})\e^{is_{a}\left(\la\tilde{\mu}\,\lambda_{a}\ra+\tilde{\chi}_{r}\eta_{a}^{r}\right)}J\cdot T_a\, ,
\end{align}
where $\bar{\delta}(z) = \dbar (1/2\pi i z)$ for complex $z$.
Note, that these vertex operators are $Q$-closed\footnote{It should be noted, that in general this theory is anomalous and has nonzero central charge, so that $Q^2 \neq 0$ \cite{ambitwistorString4d,GeyerThesis}, but one can adjust the central charge of Kac-Moody currents algebra to get $Q^2=0$ at least for lower
genus Rieman surface \cite{AmbitwistorStringsScatEqns}.} and satisfy $\{Q,\mathcal{V}_a\}=\{Q,\widetilde{\mathcal{V'}}_a\}=0$. To facilitate further comparison with Grassmannian integral representation it is convenient to introduce slightly different representation for the second vertex operator. It is obtained by a Fourier transform\footnote{Note, that in \cite{ambitwistorString4d} instead a Fourier transform for the first operator from $\eta$'s to $\tilde{\eta}$'s was performed. The Grassmann part of the
delta function is defined as usual
$\delta^{0|\mathcal{N}}(\tilde{\eta})=\prod_{r=1}^{\mathcal{N}}\tilde{\eta}_r$.} of the $\eta$'s into $\tilde{\eta}$'s:
\begin{equation}
\widetilde{\mathcal{V}}_a= \int\frac{\d s_{a}}{s_{a}}\bar{\delta}^{2|\mathcal{N}}(\tilde{\lambda}_{a}-s_{a}\tilde{\lambda}
|\tilde{\eta}_a-s_a\tilde{\chi} )\e^{is_{a}\la\tilde{\mu}\,\lambda_{a}\ra}J\cdot T_a\, .
\end{equation}
In the case of $\mathcal{N} = 3$ these vertex operators together encode all sixteen degrees of freedom of $\mathcal{N}=4$ SYM theory. For $\mathcal{N} = 4$ on the other hand each of them contains all $\mathcal{N}=4$ SYM on-shell states. In our consideration of $\mathcal{N}=4$ SYM to obtain maximally supersymmetric superamplitudes we will use the second option and use these vertex operators interchangeably.

$\text{N}^{k-2}\text{MHV}$ superamplitudes may be then obtained for example as correlation functions of $k$ operators from dual twistor space and $n-k$ operators from twistor space \cite{ambitwistorString4d}
(here and below we omit color structures and already work with color ordered objects):
\begin{equation}\label{VertexOperatorCorrelationFunction}
A_{k,n}=\left\la \widetilde{\mathcal{V}}_1 \ldots \widetilde{\mathcal{V}}_k \mathcal{V}_{k+1}\ldots  \mathcal{V}_n\right\ra\, .
\end{equation}
Instead of computing the infinite number of contractions required by exponentials in vertex operators it is convenient to take exponentials into the action as sources
\[
\int_\Sigma  \sum_{i=1}^k  i s_i \la \tilde \mu \lambda_i\ra\bar\delta(\sigma-\sigma_i) + \sum_{p=k+1}^n is_p([\mu\,\tilde\lambda_p]+\chi\tilde{\eta}_p)\bar\delta(\sigma-\sigma_p)\,.
\]
The corresponding equations of motion from this new action are then given by
\begin{align}
\bar{\partial}_\sigma Z&=\bar{\partial}\left(\lambda,\mu,\chi\right)=\sum_{i=1}^{k}s_{i}\left(\lambda_{i},0,0\right){\bar\delta}\left(\sigma-\sigma_{i}\right)\,,  \\ \label{dbar-l}
\bar{\partial}_\sigma W&=\bar{\partial}\left(\tilde{\mu},\tilde{\lambda},\tilde{\chi}\right)=\sum_{p=k+1}^{n}s_{p}\left(0,\tilde{\lambda}_{p},\tilde{\eta}_p\right){\bar\delta}(\sigma-\sigma_{p}) \,,
\end{align}
As $(Z, W)$ fields are worldsheet spinors the solutions to the above equations are unique\footnote{There no fermion zero modes on sphere.} and given by
\begin{align}
Z(\sigma)&=\left(\lambda,\mu,\chi\right)=\sum_{i=1}^{k}\frac{s_{i}\left(\lambda_{i},0,0\right)}{\sigma-\sigma_{i}}\,, \label{ZWsolution1}
\\
W(\sigma)&=\left(\tilde{\mu},\tilde{\lambda},\tilde{\chi}\right)=\sum_{p=k+1}^{n}\frac{s_{p} \left(0,\tilde{\lambda}_{p},\tilde{\eta}_p\right)}{\sigma-\sigma_{p}} \, . \label{ZWsolution2}
\end{align}
Then the path integrals over $(Z, W)$ fields localize on the solutions (\ref{ZWsolution1})-(\ref{ZWsolution2}), while current correlator contributes Parke-Taylor factor and for the color ordered on-shell amplitude we get \cite{ambitwistorString4d}:
\begin{equation}
\begin{split}
A_{n,k}=\int  \frac{1}{\text{Vol} \,\text{GL}(2,\C)}
&\prod_{a=1}^n\frac{\d s_a\d\sigma_a}{s_a (\sigma_a-\sigma_{ a+1})}
\prod_{p=k+1}^n\bar\delta^{2}
(\lambda_p-s_p\lambda(\sigma_p)) \\
&\prod_{i=1}^k  \bar\delta^{2|\mathcal{N}}(\tilde \lambda_i -s_i\tilde\lambda (\sigma_i), \tilde{\eta}_i-s_i\tilde{\chi}(\sigma_i)) \, .
\end{split} \label{onshellStringAmp}
\end{equation}
%
%
%
%
%
%
Note, that ghosts $c$ and $v$ develop\footnote{This is easy to see with the help of Riemann-Roch theorem recalling that $\text{deg}\, T_{\Sigma} = - \text{deg}\, K_{\Sigma} = 2 g -2$, where $g$ is the genus of Riemann surface.} $n_c = 3$ (number of conformal Killing vectors on sphere) and $n_v = 1$ zero modes correspondingly, which result in the $\text{GL}(2, \C)$ quotient above. In terms of homogeneous coordinates on Riemann sphere $\sigma_\alpha=\frac1s (1,\sigma)$ the rescaled by a factor $1/s$ $W$ and $Z$ fields could be written as
\begin{equation}\label{ZdefinitonsForScatteringEquations}
Z(\sigma)= \sum_{i=1}^{k}\frac{\left(\lambda_{i},0,0\right)}{(\sigma\, \sigma_{i})}
\, , \qquad
W(\sigma)= \sum_{p=k+1}^{n}\frac{(0,\tilde{\lambda}_{p},\tilde{\eta}_p) }{(\sigma \,\sigma_{p})}\, ,
\end{equation}
where $(i \,j)=\sigma_{i\alpha}\sigma_j^\alpha$. Then the final formula for the above amplitude takes the form \cite{ambitwistorString4d}:
\begin{equation}
\begin{split}
A_{n,k}=\int  \frac1{\text{Vol} \,\text{GL} (2,\C)} &\prod_{a=1}^n\frac {\d^2 \sigma_a}{(a\, a+1)}  \, \prod_{p=k+1}^n\bar\delta^{2} (\lambda_p-\lambda(\sigma_p))  \\
&
\prod_{i=1}^k  \bar\delta^{2|\mathcal{N}}(\tilde \lambda_i - \tilde\lambda(\sigma_i),
\tilde{\eta}_i-\tilde{\chi}(\sigma_i))\, .
\end{split} \label{onshellStringAmpCovariant}
\end{equation}
The scattering equations are then follow from the support of the delta functions
\begin{equation}\label{ScatteringEquations}
k_a\cdot P(\sigma_a)=\lambda_a^\alpha\tilde\lambda_a^{\dot\alpha} P_{\alpha\dot\alpha}(\sigma_a)=\lambda_a^\alpha\tilde\lambda_a^{\dot\alpha}  \lambda_\alpha(\sigma_a)\tilde\lambda_{\dot\alpha}(\sigma_a) = 0\,.
\end{equation}
It is important to note that the exact form of scattering equations themselves and scattering equations representations for amplitudes depends on which particle vertex operators were taken as $\widetilde{\mathcal{V}}_i$ and which as $\mathcal{V}_i$ in (\ref{VertexOperatorCorrelationFunction}). So, we have several equivalent representations for $A_{k,n}$. Their existence as we will see in section \ref{GrassmanniansLinkRepresentationSec} is related to the $GL(k)$ "gauge invariance" of Grassmannian integral representation for scattering amplitudes.

\section{Generalised vertex operators}
\label{StringVertexesCorrelationFunctionsSec}
In the previous section we seen the explicit form of the worldsheet vertex operators
which corresponds to the on-shell states of the $\mathcal{N}=4$ SYM field theory.
In this section we will suggest construction of composite worldsheet operators in the ambitwistor string theory which will correspond to local and none local gauge invariant operators in $\mathcal{N}=4$ SYM.

In \cite{BrandhuberConnectedPrescription} it was conjectured that it may be possible to obtain such operators considering appropriate
terms in OPE of standard vertex operators (namely OPE for Kac-Moody currents). Also ideologically similar attempts to construct generalisation of vertex operators which
should describe off-shell states was taken in \cite{StringRegge1,StringRegge2} in the context of bosonic string theory.
We, however, found such OPE based approaches unfitting for our purpose, though we do not claim that one cannot eventually succeed considering this direction.

Instead we want to leave
worldsheet structure of operators intact and consider external kinematics as only adjustable parameters. Namely as our new generalised vertex operators $\mathcal{V}^{gen.}$ we want to consider convolution of products of vertex operators $\mathcal{V}_i$ (\ref{VertexOperatorsOnShellStates}) with
``target space wave functions'' $\psi(\{\lambda_i,\tilde{\lambda}_i,\eta_i\},...)$ - some rational functions on ambitwistor space. Here by convolution we mean the integration with respect to components of
on-shell momenta $p_i=\lambda_i\tilde{\lambda}_i$ on which vertex operators depends:
\begin{eqnarray}
\mathcal{V}^{gen.}=\int\psi(\{\lambda_i,\tilde{\lambda}_i,\eta_i\},\ldots)\prod_i\mathcal{V}_i~\frac{\d^2\vll_i\d^2\vlt_i}{\text{Vol[GL(1)]}}\d^4\vlet_i.
\end{eqnarray}
Here we will understand integration with respect $\d^2\vll_i\d^2\vlt_i$ as multidimensional contour integrals which in turn will be evaluated by residues.
We will sometimes refer to this convolution as \emph{gluing operation}.
The $\ldots$ in $\psi$ corresponds to possible dependance on parameters other then
on-shell supermomenta $\lambda_i,\tilde{\lambda}_i,\eta_i$.
To describe local and none local operators in $\mathcal{N}=4$ SYM field theory
``target space wave functions'' $\psi$ should carry appropriate quantum numbers, so it is natural to take them in the form proportional to minimal form factors - tree level form factors of corresponding operator with minimal possible number of on-shell states. Such objects can be found from general symmetry arguments in the same lines as $\mbox{MHV}_3$ and $\overline{\mbox{MHV}}_3$ amplitudes \cite{ArkaniHamedSimplestQFT} or simply by explicit evaluation\footnote{Of course, it would be highly desirable to obtain some universal geometrical description of such ``target space wave functions''. The possible direction
to obtain such description is to consider appropriate polynomial solution of classical (self-dual) equation
of motion \cite{Perturbiner1}.}.

Another heuristic argument in favour of such construction is the following: to obtain vertex operator in ordinary string theory we usually consider product of appropriate polarisation vector with
combination of worldsheet fields. For example in the case of bosonic string theory
the vertex operator $V_{gr.}$ describing graviton state is given by:
\begin{eqnarray}
	V_{gr.}=\epsilon_{\pm}^{\mu\nu}V_{\mu\nu},~V_{\mu\nu}=\sqrt{g}g^{ab}\partial_a X_{\mu}
	\partial_b X_{\nu}~\exp(ip^{\rho}X_{\rho}),
\end{eqnarray}
where $X^{\mu}$ is worldsheet fields, $p^{\rho}$ is on-shell momenta of graviton and $\epsilon_{\pm}^{\mu\nu}$ is polarisation vector of graviton corresponding to momenta $p^{\rho}$.
In suggested construction the ``target space wave functions'' $\psi$ is some kind of generalisation of polarisation vector of corresponding state, but instead of ordinary scalar product we
have integrations with respect to on-shell degrease of freedom we want to eliminate and
instead of local worldsheet operator $V_{\mu\nu}$ we are considering multilocal operators
$\prod_i\mathcal{V}_i$.

The suggested construction in principle should describe any gauge invariant operator
in $\mathcal{N}=4$ SYM field theory. Initially it was successfully tested for the Wilson line operators (reggeon state creation/annihilation operators) \cite{BorkOnishchenkoAmbitwistorReggeon}.
Here we want to give more details about derivation of this result as well as consider another new simple but important example of application of our construction. Namely we want to suggest worldsheet generalised vertex operators which should describe $\mathcal{N}=4$ SYM field theory stress-tensor supermultiplet.

\subsection{Generalised worldsheet vertex operator for $\mathcal{N}=4$ SYM field theory Wilson line operator}
Following the conjectures presented above for construction of
generalised vertex operator ${\mathcal V}^{\text{WL}}$ which should describe
field theory Wilson line operator we have to choose  ``target space wave functions'' $\psi$
proportional to minimal Willson line $A^*_{2,2+1}$. Namely we will consider
 \begin{equation}
\psi\big(\{\lambda_j,\tilde{\lambda}_j,\eta_j\}_{j=i}^{i+1},\{k,\lambda_{p},\tilde{\lambda}_{p}\}\big)=A^*_{2,2+1} (g^*, \Omega_i, \Omega_{i+1})\times\mbox{colour projectors}.
 \end{equation}
This choice of $\psi$ was first considered in \cite{BorkOnishchenkoAmbitwistorReggeon}.
Here, as was mentioned before, and in Appendix \ref{appA} we will give more detailed derivation of the results of
\cite{BorkOnishchenkoAmbitwistorReggeon}.

So proceeding this way the ambitwistor string generalised vertex operator ${\mathcal V}^{\text{WL}}$ for field theory Wilson line operator insertion could be written as
 \begin{equation}
 {\mathcal V}^{\text{WL}}_{i, i+1} = \int\prod_{j=i}^{i+1}
 \frac{\d^2\vll_j\d^2\vlt_j}{\text{Vol[GL(1)]}} \d^4\vlet_j ~A^*_{2,2+1} (g^*, \Omega_i, \Omega_{i+1}) \Big|_{\vll\to -\vll} {\mathcal V}_{i} \mathcal{V}_{i+1} \Big|_{T^a T^b\to i f^{abc} T^c\to T^c} \, ,
 \end{equation}
where the vertex is supposed to be inserted at points $\sigma_i, \sigma_{i+1}$, $c$ is the color index of off-shell gluon and we have used projection of tensor product of two adjoint on-shell gluon color representations onto off-shell gluon adjoint color representation. Also because on-shell state content of $\mathcal{V}_{i}$ and
$\widetilde{\mathcal{V}}_i$ is identical we actually consider combinations
$\mathcal{V}_{i}\mathcal{V}_{i+1}$,
$\mathcal{V}_{i}\widetilde{\mathcal{V}}_{i+1}$ and
$\widetilde{\mathcal{V}}_{i}\widetilde{\mathcal{V}}_{i+1}$ in definition
of ${\mathcal V}^{\text{WL}}_{i, i+1}$ on equal footing. The minimal off-shell amplitude $A^*_{2,2+1} (g^*, \Omega_i, \Omega_{i+1})$ is given by \cite{offshell-1leg}:
 \begin{eqnarray}
 A_{2,2+1}^*(g^*,\Omega_i, \Omega_{i+1}) &=& \frac{1}{\kappa^*}\prod_{A=1}^4\frac{\partial}{\partial\vlet_p^A} \left[\frac{
    \delta^4(k+\vll_{i}\vlt_{i}+\vll_{i+1}\vlt_{i+1})
    \delta^8(\vll_p\vlet_p+\vll_{i}\vlet_{i}+\vll_{i+1}\vlet_{i+1})
 }{\abr{p\, i}\abr{i\, i+1}\abr{i+1\, p}} \right] \nonumber \\
 &=&\frac{\delta^4(k+\vll_{i}\vlt_{i}+\vll_{i+1}\vlt_{i+1})}{\kappa^*} \frac{\delta^4\left(\tilde{\eta}_{i}\abr{p\, i+1}+\tilde{\eta}_{i+1}\abr{p\, i}\right)}
 {\abr{p\, i}\abr{i\, i+1}\abr{i+1\, p}}.
 \end{eqnarray}
 Here $p$ is the off-shell gluon direction and $\kappa^*$ was defined in Section \ref{formfactorsSec} when introducing $k_T$ decomposition of the off-shell gluon momentum $k$. It should be noted, that each of $\mathcal{V}$ operators above could be exchanged for $\widetilde{\mathcal V}$ operator, so that this vertex operator representation is not unique. Note also, that the ambitwistor string vertex operator we got is non-local which may be related to the fact that the Wilson line is non-local object itself.
 The integrations over helicity spinors $\vll_i, \vlt_i$ can be performed explicitly. The details of this computation can be found in Appendix \ref{appA}, see also \cite{FormFactorsGrassmanians}. After integrations we get (here and below we always assume the action of the projection operator $\partial^4_{\tilde{\eta}_p}$ acting on ${\mathcal V}^{\text{WL}}_{i, i+1}$ and all correlation functions containing it)
 \begin{equation}
  {\mathcal V}^{\text{WL}}_{i, i+1} = \frac{\la\xi p\ra}{\kappa^*}
  \int\frac{\d\beta_2}{\beta_2}\int\frac{\d\beta_1}{\beta_1}\frac{1}{\beta_1^2\beta_2}
  {\mathcal V}_{i} \mathcal{V}_{i+1} \Big|_{T^a T^b\to i f^{abc} T^c\to T^c} \, ,
 \end{equation}
  where
\begin{align}\label{spinorsAsFunctionsOfBeta}
 &\vll_{i} = \vlluu_{i} + \beta_2\vlluu_{i+1}\, , && \vlt_{i} =
 \beta_1\vltuu_{i}  + \frac{(1+\beta_1)}{\beta_2}\vltuu_{i+1}\, ,
 &&\vlet_{i} = -\beta_1\vleuu_{i}\, , \\
&\vll_{i+1} = \vlluu_{i+1} + \frac{(1+\beta_1)}{\beta_1\beta_2}\vlluu_{i}\, ,
&& \vlt_{i+1} = -\beta_1\vltuu_{i+1} -\beta_1\beta_2 \vltuu_{i}\, , &&\vlet_{i+1} = \beta_1\beta_2\vleuu_{i}\, .
 \end{align}
with
\begin{eqnarray}\label{kDecompsition}
    \vlluu_{i}=\lambda_p,~\vltuu_{i}=\frac{\la\xi |k}{\la\xi p\ra},~\vleuu_{i}=\vlet_p;
    ~~\vlluu_{i+1}=\lambda_{\xi},~\vltuu_{i+1}=\frac{\la p |k}{\la\xi p\ra},~\vleuu_{i+1}=0,
\end{eqnarray}
where $\lambda_{\xi} \equiv\la\xi |$ is some arbitrary spinor. It is useful to identify it
with the spinor $\lambda_{q}$ coming from helicity spinor decomposition of auxiliary vector $q$ arising in
$k_T$ decomposition of off-shell gluon momentum $k$.

The off-shell amplitude with one off-shell and $n$ on-shell legs is then given by the following ambitwistor string correlation function (let us remind the reader that we are considering already color ordered object):
\begin{equation}\label{WLineVertexOperatorCorrelationFunction}
A^*_{k,n+1} =
\left\la \widetilde{\mathcal{V}}_1 \ldots \widetilde{\mathcal{V}}_k \mathcal{V}_{k+1}\ldots  \mathcal{V}_n {\mathcal V}^{\text{WL}}_{n+1, n+2}\right\ra\, .
\end{equation}
Evaluating first ambitwistor string correlator of on-shell vertexes with the help of (\ref{onshellStringAmp}) we get
\begin{align}
A^*_{k,n+1} = &\frac{\la\xi p\ra}{\kappa^*}
\int\frac{\d\beta_2}{\beta_2}\int\frac{\d\beta_1}{\beta_1}\frac{1}{\beta_1^2\beta_2}\, \frac{1}{\text{Vol} \,\text{GL} (2,\C)} \nonumber \\ &\times
\int \prod_{a=1}^{n+2}\frac{\d s_a\d\sigma_a}{s_a (\sigma_a-\sigma_{ a+1})}
\prod_{p=k+1}^{n+2}\bar\delta^{2}
(\lambda_p-s_p\lambda(\sigma_p))
\prod_{i=1}^k  \bar\delta^{2|4}(\tilde \lambda_i -s_i\tilde\lambda (\sigma_i), \tilde{\eta}_i-s_i\tilde{\chi}(\sigma_i)) \, . \label{offshellamp1}
\end{align}
We want to remind the reader that in this formula kinematical variables
$\{\lambda_i,\tilde{\lambda}_i,\eta_i\}_{i=n+1}^{n+2}$ depends on
$\beta_1$ and $\beta_2$ according to (\ref{spinorsAsFunctionsOfBeta}) -- (\ref{kDecompsition}).

Next we want to perform integrations with respect to $\beta_1$ and $\beta_2$ to obtain
formulas similar to (\ref{onshellStringAmpCovariant}).
It is rather complicated to perform integrations over $\beta_1$ and $\beta_2$ in the
expression above due to the none linear structure of delta function arguments with respect to $\beta_1$ and $\beta_2$. So we want to linearise them first. For this purpose we are introducing unity decomposition in the form \cite{UnificationResidues}:
\begin{equation}\label{ResolitionOfIndentity}
1 = \frac{1}{\text{Vol} \,\text{GL} (k)}\int\d^{k\times (n+2)}C\, \d^{k\times k} L\, (\det L)^{n+2}
\delta^{k\times (n+2)} \left(C - L\cdot C^V [s, \sigma]\right)\, ,
\end{equation}
where the integral over $L$ matrix is an integral over $\text{GL}(k)$ linear transformations and $C^V [\sigma]$ denotes the Veronese map from $(\C^2)^{n+2}/\text{GL} (2)$ to $Gr(k, n+2)$ Grassmannian \cite{UnificationResidues} (see also \cite{BrandhuberConnectedPrescription}):
\begin{align}\label{VeroneseMap}
C^V [s, \sigma] = \left(
\begin{array}{cccc}
\vdots & \vdots & \cdots & \vdots \\
\sigma^V [s_1, \sigma_1] & \sigma^V [s_2, \sigma_2] & \cdots & \sigma^V [s_{n+2}, \sigma_{n+2}] \\
\vdots & \vdots & \cdots & \vdots \end{array}
\right)\, ,\quad  \sigma^V [s, \sigma]\equiv \left(\begin{array}{c}
\xi \\ \xi\sigma \\ \vdots \\ \xi\sigma^{k-1}
\end{array}\right)\, ,
\end{align}
where \cite{SpradlinVolovichFromTwistorString,BrandhuberConnectedPrescription} :
\begin{align}
& \xi_i = s_i^{-1}\prod_{j=1, j\neq i}^k (\sigma_j - \sigma_i)^{-1}\, , && i\in (1,k), \\
& \xi_i = s_i\prod_{j=1}^k (\sigma_j - \sigma_i)^{-1}\, , && i\in (k+1,n+2).
\end{align}
Rearranging (\ref{offshellamp1}) we can write it as
\begin{align}
A^*_{k,n+1} = &\frac{\la\xi p\ra}{\kappa^*}
\int\frac{\d\beta_2}{\beta_2}\int\frac{\d\beta_1}{\beta_1}\frac{1}{\beta_1^2\beta_2}\, \frac{1}{\text{Vol} \,\text{GL} (k)} \nonumber \\ &\times
\int\d^{k\times (n+2)} C\, F (C)\, \delta^{k\times 2} (C\cdot\vlt)\delta^{k\times4} (C\cdot \vlet) \delta^{(n+2-k)\times 2} (C^{\perp}\cdot\vll)\, ,  \label{offshellamp2}
\end{align}
where
\begin{align}\label{F(C)GR}
F(C) = \int \frac{1}{\text{Vol} \,\text{GL} (2,\C)}\prod_{a=1}^{n+2}\frac{\d s_a\d\sigma_a}{s_a (\sigma_a-\sigma_{ a+1})} \d^{k\times k} L\,
\delta^{k\times (n+2)} \left(C - L\cdot C^V [s, \sigma]\right),
\end{align}
and
\begin{eqnarray}
    \delta^{k\times 2} (C\cdot\vlt)&\equiv&\prod_{a=1}^k\delta^{2} \left(\sum_{i=1}^n c_{ai}\vlt_i\right),~\delta^{(n+2-k)\times 2} (C^{\perp}\cdot\vll)\equiv\prod_{b=k+1}^{n+2}
    \delta^{2} \left(\sum_{j=1}^{n+2}c^{\perp}_{bj}\vll_j\right),\nonumber\\
    \delta^{k\times 4} (C\cdot \vlet)&\equiv&\prod_{a=1}^k\delta^{4} \left(\sum_{i=1}^{n+2} c_{ai}\vlet_i\right),
\end{eqnarray}
$C^{\perp}$ matrix is defined by identity $C\cdot (C^{\perp})^T=0$ and it is assumed that all matrix manipulations are performed after $GL(k)$ gauge fixing.
The delta functions above should be thought as $\delta (x) = 1/x$, so that the corresponding contour integral computes the residue at $x =0$ \cite{ArkaniHamedPhysicsFromGrassmannian}.
Note that now arguments of delta functions are linear in integration variables $c_{ai}$. Also
it is implemented that appropriate integration contour $\Gamma$ is chosen for $\d^{k\times (n+2)} C$ integration. We will label such contour as $\Gamma_{k,n+2}^{tree}$. We will
make some comments about explicit construction of $\Gamma_{k,n+2}^{tree}$ in the next section.

Next, by construction $F (C)$ contains $(k-2)\times (n-k)$ delta function factors  forcing integral over $C$'s to have Veronese form \cite{UnificationResidues}. In general $F(C)$ is rather complicated rational function of minors of $C$ matrix, see the discussion in Section \ref{GrassmanniansLinkRepresentationSec}. However, it could be shown that the choice of $F(C)$ in the form
\begin{equation}\label{F(C)}
F(C) = \frac{1}{(1\cdots k) (2\cdots k+1)\cdots (n+2\cdots k-1)}
\end{equation}
correctly reproduces the results of subsequent integration over Grassmannian ($C$ matrixes). Here we use standard  notations $(i_1\ldots i_k)$ to denote minors of $C$ matrix constructed from columns of $C$ with numbers $i_1,\ldots, i_k$\footnote{We hope there will be no confusion with previous definition $(i \,j)=\sigma_{i\alpha}\sigma_j^\alpha$ used in $d^2\sigma_a$ integrals over homogeneous coordinates on Riemann sphere.}. Because this this important step in our construction we give detailed discussion of this derivation, based on \cite{ArkaniHamedPhysicsFromGrassmannian} in the next chapter.

Now we can perform change of variables $c_{ai}$ that will simplify dependance on $\beta_1$ and $\beta_2$ in the integrand (linearise the dependance on $\beta_1$ and $\beta_2$ in the denominator), so that the integrals can be evaluated by residues. The computational details of this change of variables can be found in  Appendix \ref{appA}.

To evaluate the residues we found most convenient to use the notion of composite residue \cite{DualitySMatrix}. For that purpose let's define the set $S$ of points in $\mathbb{C}^n$, such that $S=\{z|z\in\mathbb{C}^n,~s(z)=0\}$ and $s(z)$ is some holomorphic function (in our case polynomial). Next, consider $n$ - form $\omega=h(z)/s(z)dz$, where $dz=dz_1\wedge \ldots dz_n$ and $h(z)$ is some other holomorphic function (in our case it is some rational function), and  define $(n-1)$ - form:
\begin{eqnarray}
res_{j}[\omega]=(-1)^{j-1}\left(\frac{h(z)}{\partial_{z_j}s(z)}\right)\Big{|}_{S}dz_{[j]},
\end{eqnarray}
with $dz_{[j]}=dz_1\wedge\ldots\wedge dz_{j-1}\wedge dz_{j+1}\wedge\ldots\wedge dz_n$. Using this definition iteratively we may define $(n-m)$ - forms as
\begin{eqnarray}
res^m[\omega]=res_{m}\circ \ldots \circ res_{1}[\omega].
\end{eqnarray}
These forms are also known as composite residue forms. Considering our integral (\ref{offshellamp2}) as such
residue form
\begin{eqnarray}\label{IntOffShellResForm}
\omega = \frac{\la\xi p\ra}{\kappa^*}
\int\frac{\d\beta_1\wedge\beta_2}{\beta_1~\beta_2}
\int\frac{\d^{k\times (n+2)} C\,}{\text{Vol} \,\text{GL} (k)} \frac{\delta^{k\times 2} \left( C
    \cdot \vltuu \right)
    \delta^{k\times 4} \left(C \cdot \vleuu \right)
    \delta^{(n+2-k)\times 2} \left(C^{\perp} \cdot \vlluu \right)}
{(1\cdots k)\ldots(n-k+2\cdots \bold{n+1})\ldots (\bold{n+2}\cdots k-1)},\nonumber\\  \label{offshellamp3}
\end{eqnarray}
where we used notations for minors of $C$ matrix
\begin{align}
(\bold{n+2}\, 1\cdots k-1) &= (1+\beta_1)(n+1\, 1\cdots k-1) + \beta_1\beta_2(n+2\, 1\cdots k-1)\, , \\
(n-k+2\cdots \bold{n+1}) &= (n-k+2\cdots n+1) + \beta_2 (n-k+2\cdots n\, n+2)\, ,
\end{align}
and for kinematical variables
\begin{align}\label{SpinorsInDeformadGrassmannian}
&\vlluu_i = \vll_i, & i = 1,\ldots  n& , &\vlluu_{n+1} &= \lambda_p ,
&\vlluu_{n+2} &= \xi \nonumber \\
&\vltuu_i = \vlt_i, & i = 1,\ldots  n& ,
&\vltuu_{n+1} &= \frac{\la\xi |k}{\la\xi p\ra},
&\vltuu_{n+2} &= - \frac{\la p |k}{\la\xi p\ra} , \nonumber \\
&\vleuu_i = \vlet_i,  & i = 1,\ldots  n& , &\vleuu_{n+1} &= \vlet_p ,
&\vleuu_{n+2} &= 0 . \nonumber \\
\end{align}
we can take residue as $res_{\beta_1=-1}\circ res_{\beta_2=0}$. This will give us the
following result
\begin{eqnarray}
res_{\beta_1=-1}\circ res_{\beta_2=0}[\omega]=A^*_{k,n+1},
\end{eqnarray}
or in more explicitly
\begin{eqnarray}\label{GrassmannianRepForWLFFacor}
A^*_{k,n+1} = \int_{\Gamma_{k,n+2}^{tree}}\frac{d^{k\times
(n+2)}C}{\text{Vol}[GL (k)]}Reg. \frac{\delta^{k\times 2} \left( C
    \cdot \vltuu \right)
    \delta^{k\times 4} \left(C \cdot \vleuu \right)
    \delta^{(n+2-k)\times 2} \left(C^{\perp} \cdot \vlluu \right)}{(1 \cdots k)\cdots (n+1 \cdots k-2) (n+2 \; 1\cdots k-1)}, \label{offshellAmpGrassmannian1}
\end{eqnarray}
where\footnote{$Reg.$ notation for this combination of minors is chosen because
such insertion regulates behaviour of $A^*_{k,n+1}$ with respect to the soft holomorphic limit
in kinematical variables with labels $n+1$ or $n+2$ \cite{offshell-1leg}.}
\begin{eqnarray}
    Reg.=\frac{\la\xi p\ra}{\kappa^{*}}\frac{(n+2 \; 1\cdots k-1)}{(n+1 \; 1\cdots k-1)}.
\end{eqnarray}
This expression is in a complete agreement with our previous derivation \cite{offshell-1leg}.
We will make some comments on the choice of integration contour $\Gamma_{k,n+2}^{tree}$ in Section \ref{GrassmanniansLinkRepresentationSec}.

Using (\ref{offshellAmpGrassmannian1}) as starting point we can performing the inverse operation - that is taking partial integrations and reducing integral in (\ref{offshellAmpGrassmannian}) to the integral over $Gr(2, n+2)$ Grassmannian\footnote{The $Gr(2, n+2)$ Grassmannian is embedded into $Gr(k, n+2)$ Grassmannian again with the help of Veronese map, see for example \cite{BrandhuberConnectedPrescription}.}:
\begin{align}\label{RSVforWLineFF}
A^*_{k,n+1} =\int
\prod_{a=1}^{n+2}\frac {\d^2 \sigma_a}{(a\, a+1)}\frac{Reg.^V}{\text{Vol} \,\text{GL} (2,\C)} \prod_{p=k+1}^{n+2}\bar\delta^{2} (\vlluu_p-\vlluu(\sigma_p))
\prod_{i=1}^k  \bar\delta^{2|4}(\vltuu_i - \vltuu(\sigma_i),
\vletuu_i-\tilde{\underline{\underline{\chi}}}(\sigma_i))\, ,
\end{align}
where $Reg.^V$ factor is given by
\begin{equation}\label{Reg.OneOnShellGluonRVA}
Reg .^V = \frac{\la\xi p\ra}{\kappa^*}\frac{(k\, n+1)}{(k\, n+2)}
\end{equation}
and doubly underlined functions are defined as
\begin{equation}\label{LambdaSigmaWLineFF}
\left(\vlluu,\underline{\underline{\mu}},\underline{\underline{\chi}}\right) = \sum_{i=1}^{k}\frac{\left(\vlluu_{i},0,0\right)}{(\sigma\, \sigma_{i})}
\, ,\qquad
\left(\tilde{\underline{\underline{\mu}}},\vltuu,\tilde{\underline{\underline{\chi}}}\right)= \sum_{p=k+1}^{n+2}\frac{(0,\vltuu_{p},\vletuu_p) }{(\sigma \,\sigma_{p})}\, .
\end{equation}
This should be equivalent to taking integrals over $\beta_1$ and $\beta_2$ directly in (\ref{offshellamp1}). (\ref{RSVforWLineFF}) can also be considered as RSV (scattering equation)
representation for Willson line form factor. Scattering equations in this case  are given by
(\ref{ScatteringEquations}) where $\lambda(\sigma)$'s and $\tilde{\lambda}(\sigma)$'s are taken from (\ref{LambdaSigmaWLineFF}).

The result for the case of amplitudes with multiple off-shell legs $A^*_{k,m+n}$ could be obtained along the same lines as previous discussion.
In the case with first $m$ particles on-shell and last $n$ being off-shell, making identical assumptions as in $n=1$ case about the form of $F(C)$ function, we would get for $A^*_{k,m+n}$:
\begin{eqnarray}\label{GrassmannianIntegralForMultipleOffShellGluons}
A^*_{k,m+n}&=&
\int_{\Gamma_{k,m+2n}^{tree}}\frac{d^{k\times (m+2n)}C}{\text{Vol}[GL (k)]}Reg.(m+1,\ldots,m+n)\times\nonumber\\
&\times&\frac{\delta^{k\times 2} \left( C
\cdot \vltuu \right)
\delta^{k\times 4} \left(C \cdot \vleuu \right)
\delta^{(m+2n-k)\times 2} \left(C^{\perp} \cdot \vlluu \right)}{(1 \cdots k)\cdots (m\cdots m+k-1) (m+1\cdots m+k)\cdots(m+2n \cdots k-1)},\nonumber\\
\end{eqnarray}
where the external kinematical variables are chosen as
\begin{align}\label{SpinorsInDeformadGrassmannianMultipleOffShellMomenta}
&\vlluu_i = \vll_i, & i = 1,\ldots  m& ,
&\vlluu_{m+2j-1} &= \lambda_{p_j} ,
&\vlluu_{m+2j} &= \xi_j, & j = 1,\ldots  n, \nonumber \\
&\vltuu_i = \vlt_i, & i = 1,\ldots  m& ,
&\vltuu_{m+2j-1} &= \frac{\la\xi_j |k_{m+j}}{\la\xi_j p_j\ra},
&\vltuu_{m+2j} &= - \frac{\la p_j |k_{m+j}}{\la\xi_j p_j\ra} , & j = 1,\ldots  n, \nonumber \\
&\vleuu_i = \vlet_i,  & i = 1,\ldots  m& , &\vleuu_{m+2j-1} &= \vlet_{p_j} ,
&\vleuu_{m+2j} &= 0 , & j = 1,\ldots  n. \nonumber \\
\end{align}
and $Reg.(m+1,\ldots,m+n)$ function is given by the products of ratios of minors of $C$ matrix:
\begin{eqnarray}\label{RegFunctionNew}
&&Reg.(m+1,\ldots,m+n) = \prod_{j=1}^n Reg(j+m),\nonumber\\
&&Reg.(j+m) = \frac{\la\xi_j p_j\ra}{\kappa^{*}_j}\frac{(2j+m~~2j+1+m\cdots 2j+k-1+m)}{(2j-1+m~~2j+1+m\cdots 2j+k-1+m)}.
\end{eqnarray}
This result also coincides with the Grassmannian integral representation for $A^*_{k,m+n}$ first conjectured in \cite{offshell-multiplelegs}.
Also note that if the number of ${\mathcal V}^{\text{WL}}_{i, i+1}$ operators is grater
than $\widetilde{\mathcal{V}}_i$, which is given by $k$, than the correlation function is equal to 0. $\widetilde{\mathcal{V}}_i$ from which ${\mathcal V}^{\text{WL}}_{i, i+1}$ may be constructed also are taken into account. It is also possible to rewrite this result in RSV (scattering equation) form:
\begin{eqnarray}\label{MultipleOff-shellGluonsAmplRVS}
A^*_{k,m+n}&=&\int
\prod_{a=1}^{n+2}\frac {\d^2 \sigma_a}{(a\, a+1)}\frac{Reg.^V(m+1,\ldots , m+n)}{\text{Vol} \,\text{GL} (2,\C)} \prod_{p=k+1}^{m + 2 n}\bar\delta^{2} (\vlluu_p-\vlluu(\sigma_p))
\nonumber\\
&\times&\prod_{i=1}^k  \bar\delta^{2|4}(\vltuu_i - \vltuu(\sigma_i),
\vletuu_i-\tilde{\underline{\underline{\chi}}}(\sigma_i))\, ,
\end{eqnarray}
where
\begin{gather}\label{Reg.MultipleOnShellGluonRVA}
Reg.^V (m+1,\ldots , m+n) = \prod_{j=1}^{n} Reg.^V (j+m)\, ,\quad
Reg.^V (j+m) = \frac{\la\xi_j p_j\ra}{\kappa_j^*}\frac{(k\, 2j-1+m)}{(k\, 2j+m)}\, ,
\end{gather}
and doubly underlined functions of $\sigma$ as before are defined in (\ref{LambdaSigmaWLineFF}). Note that now doubly underlined kinematical variables
in (\ref{LambdaSigmaWLineFF}) now should be taken from (\ref{SpinorsInDeformadGrassmannianMultipleOffShellMomenta}).
At the end, we want to stress that the explicit form of (\ref{Reg.OneOnShellGluonRVA}) and (\ref{Reg.MultipleOnShellGluonRVA}) is not unique and is in fact related to the $GL(k)$ gauge choice in (\ref{GrassmannianRepForWLFFacor}) and (\ref{GrassmannianIntegralForMultipleOffShellGluons}).

\subsection{Generalised worldsheet vertex operators for $\mathcal{N}=4$ SYM
field theory Stress-tensor supermultiplet operators insertion}
Let's now consider different choice of ``target space wave functions'' $\psi$.
Namely let's choose $\psi$ as:
\begin{equation}\label{VoperForST}
\psi\big(\{\lambda_j,\tilde{\lambda}_j,\eta_j\}_{j=i}^{i+1},\{q,\gamma^-\}\big)=F_{2,2} (\Omega_{n}, \Omega_{n+1};\mathcal{T})\times\mbox{colour projectors}.
\end{equation}
Here $F_{2,2}$ is minimal form factor of operators from $\mathcal{N}=4$ SYM stress tensor supermultiplet\footnote{More acurately its chiral truncation \cite{HarmonyofFF_Brandhuber}.}.
This choice of ``target space wave function'' should correspond to the worldsheet generalised vertex operator ${\mathcal V}^{\text{ST}}$ which should describe the insertions of operators from $\mathcal{N}=4$ SYM Stress-tensor supermultiplet in the on-shell amplitude i.e. corresponding form factor:
\begin{equation}
{\mathcal V}^{\text{ST}}_{i, i+1} = \int\prod_{j=i}^{i+1}
\frac{\d^2\vll_j\d^2\vlt_j}{\text{Vol[GL(1)]}} \d^4\vlet_j F_{2,2} (\Omega_{i}, \Omega_{i+1};\mathcal{T}) \Big|_{\vll\to -\vll} {\mathcal V}_{i} \mathcal{V}_{i+1} \Big|_{T^a T^b\to \delta^{ab}\to 1} \, ,
\end{equation}
where, as in the previous case, the vertex is supposed to be inserted at points $\sigma_i, \sigma_{i+1}$ and we have used projection of tensor product of two adjoint on-shell state color representations onto singlet color representation. Note that initial correlation function
of vertex operators (\ref{VertexOperatorCorrelationFunction})
is colored object. The singlet projection considered here will effectively lead
to the situation when on the level of colour ordered objects we will have to consider
all possible positions\footnote{Let us remind that we consider combinations
$\mathcal{V}_{i}\mathcal{V}_{i+1}$,
$\mathcal{V}_{i}\widetilde{\mathcal{V}}_{i+1}$ and
$\widetilde{\mathcal{V}}_{i}\widetilde{\mathcal{V}}_{i+1}$ in definition
of ${\mathcal V}^{\text{ST}}_{i, i+1}$ on equal footing.} of ${\mathcal V}^{\text{ST}}_{i, i+1}$ (``gluing positions'') starting from $i=1$ up
to $i=n+1$.
The minimal form factor $F_{2,2}(\Omega_{i}, \Omega_{i+1};\mathcal{T})$ itself is given by \cite{FormFactorsGrassmanians}:
\begin{align}
F_{2,2}(\Omega_{i}, \Omega_{i+1};\mathcal{T}) = \delta^2 (\vltu_{i})\delta^4 (\vleu_{i})
\delta^2 (\vltu_{i+1})\delta^4 (\vleu_{i+1}) \label{minimalSETformfactor}
\end{align}
with ($q$ and $\gamma^-$ are the operator's momentum and supermomentum correspondingly)
\begin{align}
&\vltu_{i} = \vlt_{i} - \frac{\la i+1|q}{\la i+1\, i\ra}\, , \quad
&\vleu^-_{i} = \vlet^-_{i} - \frac{\la i+1|\gamma^-}{\la i+1\, i\ra}\, ,
\quad &\vleu^+_{i} = \vlet^+_{i}\, ,  \\
&\vltu_{i+1} = \vlt_{i+1} - \frac{\la i|q}{\la i\, i+1\ra}\, , \quad
&\vleu^-_{i+1} = \vlet^-_{i+1} - \frac{\la i|\gamma^-}{\la i\, i+1\ra}\, ,
\quad &\vleu^+_{i+1} = \vlet^+_{i+1}\, .
\end{align}
Integrating over helicity spinors $\vll_i, \vlt_i$ we get
 \begin{equation}
 {\mathcal V}^{\text{ST}}_{i, i+1} =-\la\xi_A\xi_B\ra^2
 \int\d\beta_1\int\d\beta_2\,
 {\mathcal V}_{i} \mathcal{V}_{i+1} \Big|_{T^a T^b\to \delta^{a b}\to 1} \, ,
 \end{equation}
where
 \begin{align}\label{FormFKinematicalBetaDependance1}
&\vll_{i} = \xi_A - \beta_1\xi_B\, ,\quad &\vlt_{i} = \frac{1}{\beta_1\beta_2-1}\frac{\la\xi_B|q}{\la\xi_B\xi_A\ra} + \frac{\beta_2}{\beta_1\beta_2 - 1}\frac{\la\xi_A |q}{\la\xi_A\xi_B\ra}  \\
&\vll_{i+1} = \xi_B - \beta_2\xi_A\, ,\quad &\vlt_{i+1} = \frac{1}{\beta_1\beta_2-1}\frac{\la\xi_A|q}{\la\xi_A\xi_B\ra} + \frac{\beta_1}{\beta_1\beta_2 - 1}\frac{\la\xi_B |q}{\la\xi_B\xi_A\ra}
 \end{align}
and
\begin{align}
&\vlet^-_{i} = \frac{1}{\beta_1\beta_2 - 1}\frac{\la\xi_B|\gamma^-}{\la\xi_B\xi_A\ra} + \frac{\beta_2}{\beta_1\beta_2 - 1}\frac{\la\xi_A|\gamma^-}{\la\xi_A\xi_B\ra},\quad  &\vlet^+_{i} = 0 \\
&\vlet^-_{i+1} = \frac{1}{\beta_1\beta_2 - 1}\frac{\la\xi_A|\gamma^-}{\la\xi_A\xi_B\ra} + \frac{\beta_1}{\beta_1\beta_2 - 1}\frac{\la\xi_B|\gamma^-}{\la\xi_B\xi_A\ra},\quad  &\vlet^+_{i+1} = 0\label{FormFKinematicalBetaDependance2}
\end{align}
Evaluation of string correlation function with stress-tensor vertex operator insertion
(here we are considering colour ordered object)
\begin{equation}\label{STineVertexOperatorCorrelationFunction}
F_{k,n} =
\left\la \widetilde{\mathcal{V}}_1 \ldots \widetilde{\mathcal{V}}_k \mathcal{V}_{k+1}\ldots  \mathcal{V}_n {\mathcal V}^{\text{ST}}_{n+1, n+2}\right\ra+\text{\it other gluing positions}\, .
\end{equation}
closely follows the corresponding calculation for the case of Wilson line vertex operator insertion presented above. We also want to simplify arguments of delta functions. For that purpose we introduce unity decomposition \cite{UnificationResidues} in the form (\ref{ResolitionOfIndentity})
and use conjecture that $F(C)$ can be chosen as (\ref{F(C)}). This give us the following expression
\begin{align}
F_{k,n} &= -\la\xi_A\xi_B \ra^2
\int\d\beta_2\d\beta_1
\int\frac{\d^{k\times (n+2)} C}{\text{Vol}[GL(k)]}\, F (C)\, \delta^{k\times 2} (C\cdot\vlt)\delta^{k\times4} (C\cdot \vlet) \delta^{(n+2-k)\times 2} (C^{\perp}\cdot\vll)\,  \nonumber \\ &+\text{\it other gluing positions}\,\label{FormFactorsIntermidiate}
\end{align}
Note once more, that here in the expression above in all terms $\lambda_i,\tilde{\lambda}_i,\tilde{\eta}_i$ are functions
of $\beta_1$ and $\beta_2$ according to (\ref{FormFKinematicalBetaDependance1}) - (\ref{FormFKinematicalBetaDependance2}). In the first term explicitly written here $i=n+1$,
in the second $i=n$ e t.c.
After appropriate change of variables, which is given in Appendix \ref{appA}, we can rewrite (\ref{FormFactorsIntermidiate}) as (similar to the previous case we understand integration over $\beta_1,\beta_2$ as residue form $\omega$, though in this case it dose not bring any simplifications):
\begin{eqnarray}
\omega &=& -\la\xi_A\xi_B\ra^2  \int\frac{d\beta_1\wedge d\beta_2}{(1-\beta_1\beta_2)}\int
\frac{d^{k\times (n+2)} C}{\text{Vol}[GL(k)]}
 \delta^{k\times 2}\left( C'
\cdot \vltuu \right)
\delta^{k\times 4} \left(C'\cdot \vleuu \right)
\delta^{(n+2-k)\times 2} \left(C'^{\perp}\cdot \vlluu \right) \nonumber \\
&\times& \frac{1}{(1\cdots k)\cdots(n-k+2\cdots n\, \bold{n+1})\cdots(\bold{n+2}\, 1\cdots k-1)}+\text{\it other gluing positions}\, .\nonumber\\	
\end{eqnarray}
Here the following notations was used for minors
\begin{eqnarray}
(n-k+2\cdots n\, \bold{n+1})&=&(n-k+2\cdots n\, n+1)-\beta_1(n-k+2\cdots n\, n+2),\nonumber\\
(\bold{n+2}\, 1\cdots k-1)&=&(n+2\, 1\cdots k-1)-\beta_2(n+1\, 1\cdots k-1),
\end{eqnarray}
and for kinematical variables:
\begin{align}\label{SpinorsFormFactorGrassmannian}
&\vlluu_i = \vll_i, & i = 1,\ldots  n& , &\vlluu_{n+1} &= \xi_A ,
&\vlluu_{n+2} &= \xi_B \nonumber \\
&\vltuu_i = \vlt_i, & i = 1,\ldots  n& ,
&\vltuu_{n+1} &= -\frac{\la\xi_B |q}{\la\xi_B\xi_A\ra},
&\vltuu_{n+2} &= - \frac{\la\xi_A |q}{\la\xi_A\xi_B \ra} , \nonumber \\
&\vleuu_i^+ = \vlet_i^+,  & i = 1,\ldots  n& , &\vleuu^+_{n+1} &= 0 ,
&\vleuu^+_{n+2} &= 0 , \nonumber \\
&\vleuu^-_i = \vlet^-_i,  & i = 1,\ldots  n& , &\vleuu^-_{n+1} &= -\frac{\la\xi_B|\gamma^-}{\la\xi_B\xi_A\ra} ,
&\vleuu^-_{n+2} &=  -\frac{\la\xi_A|\gamma^-}{\la\xi_A\xi_B\ra}. \nonumber \\
\end{align}
Taking residues at $\beta_1^* = \frac{(n-k+2\cdots n\, n+1)}{(n-k+2\cdots n\, n+2)}$ and $\beta_2^* = \frac{(n+2\, 1\cdots k-1)}{(n+1\, 1\cdots k-1)}$ we reproduce the result of \cite{FormFactorsGrassmanians} (computational details can also be found in Appendix \ref{appA}.):
\begin{equation}
F_{k,n} = res_{\beta_1=\beta_1^*}\circ res_{\beta_2=\beta_1^*}[\omega],
\end{equation}
where (here we write explicitly only first term corresponding to ${\mathcal V}^{\text{ST}}_{i, i+1}$ positioned in $i=n+1$):
\begin{eqnarray}\label{GrReprFormfactorST}
F_{k,n} &=&  \int_{\Gamma^{tree,n+1}_{k,n+2}}
\frac{d^{k\times (n+2)} C}{\text{Vol}[GL(k)]}
Reg.
\frac{\delta^{k\times 2} \left( C
	\cdot \vltuu \right)
	\delta^{k\times 4} \left(C\cdot \vleuu \right)
	\delta^{(n+2-k)\times 2} \left(C^{\perp}\cdot \vlluu \right)}{(1\cdots k)(2\cdots k+1)\cdots (n+2\cdots k-1)}\,\nonumber\\
&+&\text{\it other gluing positions}\, ,
\end{eqnarray}
with\footnote{Similar to the previous case $Reg.$ insertion regulates soft holomorphic limit with respect to kinematical variables with labels $n+1$ \emph{and} $n+2$ \cite{SoftTheoremsFormFactors}.}
\begin{equation}
Reg.=\la\xi_A\xi_B\ra^2\frac{Y}{1-Y},~Y = \frac{(n-k+2\cdots n\, n+1)(n+2\, 1\cdots k-1)}{(n-k+2\cdots n\, n+2)(n+1\, 1\cdots k-1)} .
\end{equation}
Additional label in $\Gamma^{tree,n+1}_{k,n+2}$ corresponds to the fact that for each term
corresponding to different ${\mathcal V}^{\text{ST}}_{i, i+1}$ positions
integration contours should be, in general, chosen separately \cite{FormFactorsGrassmanians}.

Using obtained above expression (\ref{GrReprFormfactorST}) as in the previous discussion we can performing the inverse operation - that is taking partial integrations and reducing integral in (\ref{offshellAmpGrassmannian}) to the integral over $Gr(2, n+2)$ Grassmannian
\begin{eqnarray}\label{sc1}
F_{k,n} &=&\int
\prod_{a=1}^{n+2}\frac {\d^2 \sigma_a}{(a\, a+1)}\frac{Reg.}{\text{Vol} \,\text{GL} (2,\C)} \prod_{p=k+1}^{n+2}\bar\delta^{2} (\vlluu_p-\vlluu(\sigma_p))
\prod_{i=1}^k  \bar\delta^{2|4}(\vltuu_i - \vltuu(\sigma_i),
\vletuu_i-\tilde{\underline{\underline{\chi}}}(\sigma_i))\,\nonumber\\
&+&\text{\it other gluing positions}\, ,
\end{eqnarray}
where $Reg.$ factor is now given by \cite{BrandhuberConnectedPrescription}:
\begin{equation}
Reg . =  \la\xi_A\xi_B\ra^2\frac{Y}{1-Y} \, ,\quad Y = \prod_{j=n+2-k}^n\frac{(j\, n+1)}{(j\, n+2)}\prod_{i=1}^{k-1}\frac{(n+2\, i)}{(n+1\, i)}
\end{equation}
and doubly underlined functions are defined as in the case of Wilson line insertion:
\begin{equation}
\left(\vlluu,\underline{\underline{\mu}},\underline{\underline{\chi}}\right) = \sum_{i=1}^{k}\frac{\left(\vlluu_{i},0,0\right)}{(\sigma\, \sigma_{i})}
\, ,\qquad
\left(\tilde{\underline{\underline{\mu}}},\vltuu,\tilde{\underline{\underline{\chi}}}\right)= \sum_{p=k+1}^{n+2}\frac{(0,\vltuu_{p},\vletuu_p) }{(\sigma \,\sigma_{p})}\, .
\end{equation}
This should be equivalent to direct calculation of $\beta_{1,2}$ integrals in (\ref{WLineVertexOperatorCorrelationFunction}). This also can be considered as analog
of RSV (scattering equation) representation of form factors of stress tensor supermultiplet operator.

Let us remind the reader once more that in the formula above the term $+\, \text{\it other }\,$ $\text{\it gluing positions}\,$ denotes all other insertion positions of minimal form factor in the color ordered on-shell amplitude. Note, that the original string correlation function contains all these terms corresponding to different gluing positions from the very beginning.

At the and of this section let us make the following comment regarding results presented in the literature \cite{BrandhuberConnectedPrescription, Meidinger:2017hvm}. Scattering equation representation obtained here is different from the main result of \cite{BrandhuberConnectedPrescription} (see see 2.11 there). We want to stress that we reproduce results of \cite{FormFactorsGrassmanians} starting from vertex operator correlation function and our definition of generalised vertex operator (\ref{VoperForST}), at least if we fix ``appropriate'' integration order and will take integrals with respect to $\beta_{1,2}$ in (\ref{FormFactorsIntermidiate}) as the last one. Scattering equation representation (\ref{sc1}) also obviously coincides with results of application of Veronese map to individual terms of Grassmannian representation of 
\cite{FormFactorsGrassmanians}. We also have checked that we reproduce
answers for NMHV $n=3,4,5$ point and NNMHV $n=4$ point form factors, similar to \cite{FormFactorsGrassmanians} if we use integration ordering described above. We think that both scattering equation representations i.e. \cite{BrandhuberConnectedPrescription} and (\ref{sc1}) give in the end (after integration)
identical result and the explanation to this is that the different functions can have coinciding subset of residues. 

\section{Grassmannians, scattering equations and link representations}\label{GrassmanniansLinkRepresentationSec}
In derivation of Grassmannian representations (\ref{GrassmannianRepForWLFFacor}) and (\ref{GrReprFormfactorST}) from ambitwistor string worldsheet correlation functions (\ref{WLineVertexOperatorCorrelationFunction}) and (\ref{STineVertexOperatorCorrelationFunction}) it was crucial that we can choose $F(C)$ function in the form of (\ref{F(C)}). So
for self consistency
here we want to present detailed discussion and give arguments that such choice is
indeed possible. Our discussion will be based mostly on \cite{UnificationResidues}, so if the reader is familiar with the content of \cite{UnificationResidues} he/she can skip reading to the end of this section where explicit example is considered for the Willson line field theory operator form factor with $n+2=6$
and $k=3$.

Let's look at the case of Wilson line operator form factors and consider (\ref{offshellamp2}). For stress tensor supermultiplet operator form factors we will have similar expression but with different $\beta_{1,2}$ dependance. For fixed values of $\beta_{1,2}$ parameters the integrand of
(\ref{offshellamp2}) looks like:
\begin{align}\label{BetaIntegrand}
Int =
\int\frac{\d^{k\times (n+2)} C\,}{\text{Vol} \,\text{GL} (k)} F (C)\, \delta^{k\times 2} (C\cdot\vlt)\delta^{k\times4} (C\cdot \vlet) \delta^{(n+2-k)\times 2} (C^{\perp}\cdot\vll)\, ,
\end{align}
All dependance on $\beta_{1,2}$ is accumulated in this expression in $\{\lambda_i,\tilde{\lambda}_i,\eta_i\}_{i=n+1}^{n+2}$
and is given by (\ref{spinorsAsFunctionsOfBeta}). Here
\begin{eqnarray}\label{F(C)fromString}
F(C)&=&\int \frac{1}{\text{Vol} \,\text{GL}(2,\C)}
\prod_{b=1}^n\frac{\d s_b\d\sigma_b}{s_a (\sigma_b-\sigma_{ b+1})}
\prod_{\substack{a\in f \\ i\in g}} \delta
\left(c_{ai}-\frac{s_as_i}{\sigma_a-\sigma_i}\right),
\end{eqnarray}
with $f, g$ denoting index sets $f=1,\ldots,k$ and $g=k+1,\ldots,n+2$\footnote{This
particular form of $f$ and $g$ is related to $GL(k)$ gauge choice. Namely, $f$ contains the numbers of columns constituting unity matrix. The different choices of $f$ and $g$ sets with given total numbers of elements $\# f=k$, $\# g = n-k+2$ in each set correspond to different gauge choices and also to different rearrangements of $\mathcal{V}_a$ and $\widetilde{\mathcal{V}}_a$ vertex operator among themselves in correlation function. All gauge should lead to the same result.}. Note that $F(C)$
is completely kinematically independent and will have the same form also for the case
of stress tensor supermultiplet form factors.
In fact $Int$ is identical to the RSV representation of $n+2$ point $\mbox{N}^{k-2}\mbox{MHV}$ amplitude: $Int=A_{k,n+2}$ (it is implemented that appropriate integration contour $\Gamma$ is chosen) with appropriately chosen kinematics. So let's forget for now about $\beta_{1,2}$ integrals completely
and concentrate on the RSV representation of $A_{k,n+2}$.
Let's transform the expression for (\ref{BetaIntegrand}) into more suitable for our purpose form.
For that it is convenient  to  rearrange delta functions of kinematical constraints
in the form \cite{DualitySMatrix}:
\begin{eqnarray}\label{FromCtoTau1}
&& \delta^{k\times 2} (C\cdot\vlt)\delta^{(n+2-k)\times 2} (C^{\perp}\cdot\vll)=
\delta^4\left(\sum_{j=1}^{n+2}\lambda_j\tilde{\lambda}_j\right)~J(\lambda,\tilde{\lambda})\times\nonumber\\
&\times&
\int d^{(k-2)(n-k)}\tau_A ~\prod_{\substack{a\in f \\ i\in g}}\delta\left(c_{ai}-c_{ai}(\tau|kin.)\right),
\end{eqnarray}
where $J(\lambda,\tilde{\lambda})$ is the Jacobian of transformation and $c_{ai}(\tau|kin.)$ is a general solution of underdetermined system of linear equations \cite{DualitySMatrix,GoddardGluonTreeAmplitudesinOpenTwistorString}
\begin{eqnarray}\label{underdeterminedSystemOfEquations}
c_{ai}\lambda_a&=&-\lambda_i,\nonumber\\
c_{ai}\tilde{\lambda}_i&=&-\tilde{\lambda}_a,
\end{eqnarray}
with $a\in f,~i\in g$. The solution depends on external kinematical data $\lambda_i,\tilde{\lambda}_i$ as well as on the arbitrary $(k-2)(n-k)$ parameters $\tau_A$. The explicit form of $c_{ai}(\tau|kin.)$ for general $n$ and $k$ can be found in \cite{GoddardGluonTreeAmplitudesinOpenTwistorString}. For example \cite{DualitySMatrix}, for $n+2=6$, $k=3$ and $f \in (1,3,5)$ and $g \in (2,4,6)$ we have $c_{ai}(\tau|kin.)=c_{ai}^*+\epsilon_{aa_1a_2}
\epsilon_{ii_1i_2}\langle a_1a_2\rangle[i_1i_2]\tau$, where $c_{ai}^*$ is some particular solution of (\ref{underdeterminedSystemOfEquations}).
Using the representation (\ref{FromCtoTau1}) we can remove integration over $d^{k\times (n+2)}C/\text{Vol}[GL (k)]$. Next, let's for simplicity fix helicities of external particles in such a way that Grassmann
delta functions  $\delta^{k\times 4}$ go to 1\footnote{This is always possible for appropriate $GL(k)$ gauge and external state choices. For example, for $n+2=6,~k=3$ and $f=2,4,6$, $g=1,3,5$ the appropriate choice of the external particles helicities will be $(+-+-+-)$ \cite{DualitySMatrix}.}. All these manipulations
reduce our initial expression (\ref{BetaIntegrand}) to
\begin{eqnarray}
\label{AmpTauIntegral}
A_{k,n+2}&=&\delta^4\left(\sum_{j=1}^{n+2}\lambda_j\tilde{\lambda}_j\right)
~J(\lambda,\tilde{\lambda})~ \int_{\Gamma} d^{(k-2)(n-k)}\tau_A
~F(C)\big{|}_{c_{ai}\mapsto c_{ai}(\tau|kin.)},
\end{eqnarray}
with the appropriate choice of integration contour $\Gamma$.

Now one can try to evaluate the function $F(C)$ for general values of $n$ and $k$ in terms of matrix elements
$c_{ai}$ \cite{GoddardGluonTreeAmplitudesinOpenTwistorString}. It is a rather complicated expression. The
most studied case is $k=3$ \cite{UnificationResidues,SpradlinVolovichFromTwistorString,VolovichAGrassmannianEtudeinNMHVMinors} and it is believed that for $k > 3$ the behavior will be essentially the same as in $k=3$ case \cite{UnificationResidues}. Then let us also concentrate on the $k=3$ case as representative, yet simple enough example. In this case we can rewrite $F(C)$ function in terms of minors of $C$ matrix and get \cite{UnificationResidues,VolovichAGrassmannianEtudeinNMHVMinors}
\begin{eqnarray}\label{F(C)stringk=3}
F^{k=3}(C)=H(C)\frac{1}{S_6\ldots S_{n+2}},
~H(C)=\frac{\prod_{j=6}^{n+1}(12j)(23j-1)\prod_{i=5}^{n+1}(13i)}{(n+1n+21)(123)(234)},
\end{eqnarray}
and ($j=6,\ldots,n+2$)
\begin{eqnarray}
S_j=(j-2j-1j)(j12)(23j-2)(j-113)-(j-1j1)(123)(3j-2j-1)(j2j-2).\nonumber\\
\end{eqnarray}
Note that it is different from our choice (\ref{F(C)}).

Let us summarise what we have learned so far. We have explicitly evaluated integral (\ref{F(C)fromString}) for $k=3$
and found that the result of evaluation is naively different from what we have conjectured.
Presumably $k >3$ will be no better.
To see how this contradiction resolves let us consider consider representation of the amplitude $A_{k,n+2}$ in terms of the integral over Grassmannian $Gr(k,n+2)$
\begin{eqnarray}
A_{k,n+2} = \int_{\Gamma_{k,n+2}^{tree}}\frac{d^{k\times
		(n+2)}C}{\text{Vol}[GL (k)]}\frac{\delta^{k\times 2} \left( C
	\cdot \vltuu \right)
	\delta^{k\times 4} \left(C \cdot \vleuu \right)
	\delta^{(n+2-k)\times 2} \left(C^{\perp} \cdot \vlluu \right)}{(1 \cdots k)\cdots (n+1 \cdots k-2) (n+2 \; 1\cdots k-1)}. \label{offshellAmpGrassmannian}
\end{eqnarray}
Using the same as before manipulations (namely (\ref{FromCtoTau1}) and (\ref{underdeterminedSystemOfEquations})) we will arrive at a similar expression (\ref{AmpTauIntegral})  \cite{UnificationResidues}, but with different form of $F(C)$ function, which we will denote now as $F_{Gr}(C)$:
\begin{equation}
A_{n+2,k}=\delta^4\left(\sum_{j=1}^{n+2}\lambda_j\tilde{\lambda}_j\right)~J(\lambda,\tilde{\lambda})~
\int_{\Gamma_{k,n+2}^{tree}} d^{(k-2)(n-k)}\tau_A
~F_{Gr}(C)\big{|}_{c_{ai}\mapsto c_{ai}(\tau|kin.)},
\end{equation}
where for $k=3$
\begin{eqnarray}
F^{k=3}_{Gr}(C)=\tilde{H}(C)\frac{1}{\tilde{S}_6\ldots
	\tilde{S}_{n+2}},
~\tilde{H}(C)=\frac{\prod_{j=6}^{n+1}(12j)(23j-1)}{(n+1n+21)(123)(234)},
\end{eqnarray}
and
\begin{eqnarray}
\tilde{S}_j=(j-2j-1j)(j12)(23j-2),~j=6,\ldots,n+2.
\end{eqnarray}
The $F^{k=3}_{Gr}(C)$ function is given by essentially rearranged cyclic factor \cite{UnificationResidues}:
\begin{eqnarray}\label{F(C)Gr}
F^{k=3}_{Gr}(C)=\frac{1}{(123)(234)\ldots(n+212)}
\end{eqnarray}
Note that now this form of $F(C)$ corresponds to our choice (\ref{F(C)}).

As side note let's point out that one can consider $S_j$ or $\tilde{S}_j$ function as the explicit construction of the map $\bold{S}=(\tilde{S}_6,\ldots,\tilde{S}_{n+2})$, $\bold{S}:\mathbb{C}^{(n-3)} \mapsto \mathbb{C}^{(n-3)}$, which zeros
determine the integration contour\footnote{It is also important to mention that for $k>3$ analogs of $(\tilde{S}_6,\ldots,\tilde{S}_{n+2})$
maps $\bold{S}:\mathbb{C}^{(k-2)(n-k)} \mapsto \mathbb{C}^{(k-2)(n-k)}$ may be also constructed \cite{GoddardGluonTreeAmplitudesinOpenTwistorString,Talesof1001Gluons} and thus the explicit form of $\Gamma_{k,n+2}^{tree}$ integration contours is known.} $\Gamma=\Gamma_{3,n+2}^{tree}$.

So the natural question is how these different expressions ($F(C)$ and $F_{Gr}(C)$) can provide us with the representation of the same object?  The answer to this question and
also resolution of our contradiction was given in \cite{UnificationResidues,VolovichAGrassmannianEtudeinNMHVMinors}. It turns out, that there actually exists a family of functions $F^{k=3}(C|t_6,\ldots,t_{n+2})$ depending on parameters $t_6,\ldots,t_{n+2}$, such that:
\begin{eqnarray}
&&F^{k=3}(C|t_6,\ldots,t_{n+2})=H(C)\frac{1}{S_6(t_6)\ldots
	S_{n+2}(t_{n+2})},\nonumber\\
&&H(C)=\frac{\prod_{j=6}^{n+1}(12j)(23j-1)\prod_{i=5}^{n+1}(13i)}{(n+1n+21)(123)(345)},
\end{eqnarray}
with ($j=6,\ldots,n+2$)
\begin{eqnarray}\label{S(t)conic}
S_j(t)=(j-2j-1j)(j12)(23j-2)(j-113)-t_j(j-1j1)(123)(3j-2j-1)(j2j-2),\nonumber\\
\end{eqnarray}
so that the result of evaluating by residues at zeros of
$\textbf{S}(t)=(S_6(t_6),\ldots,S_{n+2}(t_{n+2}))$ map
the integral (\ref{AmpTauIntegral}) is $t_j$ independent \cite{UnificationResidues}:
\begin{eqnarray}
\partial_{t_j}\int_{\textbf{S}(t)=0} d^{(n-k)}\tau_A
~F^{k=3}(C|t_6,\ldots,t_{n+2})\big{|}_{c_{ai}\mapsto
	c_{ai}(\tau|kin.)}=0,~\mbox{for}~j=6,\ldots,n+2.
\end{eqnarray}
The case $t_j=0$ corresponds to the representation of amplitude obtained from Grassmannian integral representation, while the case $t_j=1$ corresponds to the representation obtained from scattering equations representation:
\begin{eqnarray}
F^{k=3}(C|1,\ldots,1)=F^{k=3}(C),~\mbox{and}~
F^{k=3}(C|0,\ldots,0)=F^{k=3}_{Gr}(C).
\end{eqnarray}
The obtained relation thus supports the assertion, that Grassmannian integral representation has stringy origin.

As an illustration let's consider simplest case $k=3$, $n+2=6$. In this case we have integral over single
complex parameter $\tau$ (it is assumed that in all minors the replacement $c_{ai}\mapsto c_{ai}(\tau|kin.)$ was performed):
\begin{eqnarray}
A_{6,3}&=&\int_{S(t)=0} d\tau\frac{(135)}{(123)(345)(561)}\frac{1}{S(t)},\nonumber\\
S(t)&=&t(123)(345)(561)(246)-(234)(456)(612)(351)\, ,
\end{eqnarray}
where minors $(123)$, $(345)$, $(561)$ and $S(t)$ are liner function of $\tau$. According to Cauchy theorem the different residues are related with each other as
\begin{eqnarray}
\{S(t)\}=-\{(123)\}-\{(345)\}-\{(561)\} .
\end{eqnarray}
Here $\{\ldots\}$ denotes the integral residue at the corresponding pole. Note, that for $(123)=0$, $(345)=0$ or $(561)=0$ the term in $S(t)$ proportional to $t$
vanishes and as a consequence we have
\begin{eqnarray}
\partial_t\{(123)\}=\partial_t\{(345)\}=\partial_t\{(561)\}=0,
\end{eqnarray}
So, in the computation of the above integral we can put
$S(t)$ to $S(0)$ and get
\begin{eqnarray}
\frac{(135)}{(123)(345)(561)}\frac{1}{S(0)}=\frac{1}{(123)\ldots(612)}.
\end{eqnarray}
In the case $n+2 >6$ the situation is similar, but now one must deal with multiple integrations over complex variables and use global residue theorem \cite{UnificationResidues,VolovichAGrassmannianEtudeinNMHVMinors}. The explicit computations were also preformed for the $k=4$ case in \cite{UnificationResidues} and it is believed that one can use $F(C)$ function in the form of
\begin{equation}\label{F(C)fromGrassmannian}
F(C) =F_{Gr}(C)= \frac{1}{(1\cdots k) (2\cdots k+1)\cdots (n+2\cdots k-1)}.
\end{equation}
for general values of $n$ and $k$, which is why we also used $F(C)$ function in the form of (\ref{F(C)}) in our considerations in the previous chapter. However, as far as we know there is no general proof of this assertion.

Let's once more stress that in all considerations above we never used explicit form of kinematical dependance of (i.e. explicit form of solution of \ref{underdeterminedSystemOfEquations}) minors of matrix $C$. So presented above construction will be valid not only for the $A_{k,n+2}$ amplitude but also for the integrand
of (\ref{offshellamp2}) and in the analogous expression for the form factors of operators
from stress tensor supermultiplet, where, in both cases, some $\lambda$'s and $\eta$'s are rational functions of
$\beta_{1,2}$ parameters. This is why we can replace $F(C)$ with $F_{Gr}(C)$ according to (\ref{F(C)}).

In the end of this section, as an example, let us consider the simple none trivial case of $n+2=6$, $k=3$ and check that we indeed get the same result independent of whether we faithfully use (\ref{F(C)fromString}) or replace it with (\ref{F(C)fromGrassmannian}) as  $F(C)$ function in (\ref{offshellamp2}).
Computing the string correlation function from the previous section for $A^*_{3,4+1}$ amplitude we end up with the following expression (let's stress once again that in
this formula $\lambda_i,\tilde{\lambda}_i,\tilde{\eta}_i$ for $i=5,6$ are functions of $\beta_{1,2}$ according to (\ref{spinorsAsFunctionsOfBeta})):
\begin{align}
A^*_{3,4+1} = &\frac{\la\xi p\ra}{\kappa^*}
\int\frac{\d\beta_1\wedge\d\beta_2}{\beta_1\beta_2}\frac{1}{\beta_1^2\beta_2}\, \frac{1}{\text{Vol} \,\text{GL} (3)} \nonumber \\ &\times
\int\d^{3\times 6} C\, F (C)\, \delta^{k\times 2} (C\cdot\vlt)\delta^{3\times4} (C\cdot \vlet) \delta^{(3)\times 2} (C^{\perp}\cdot\vll)\, ,  \label{offshellamp2Example}
\end{align}
where $F(C)$ is given by (\ref{F(C)stringk=3}) with $n=4$.
Performing change of variables (see also appendix A) and
evaluating composite residue at points $res_{\beta_1=-1}\circ res_{\beta_2=0}$
we end up with
\begin{eqnarray}\label{ExampleA*NMHVcontour}
A^*_{3,4+1}&=&\int_{S=0} d\tau ~\tilde{F}(C), ~\tilde{F}(C)=\frac{(135)}{(123)(345)(561)}\frac{1}{S},\nonumber\\
S&=&(123)(345)(561)(245)-(234)(456)(512)(351),
\end{eqnarray}
and all minors, according to (\ref{underdeterminedSystemOfEquations}) are functions of external kinematical data defined as (\ref{SpinorsInDeformadGrassmannian}) with $n=4$, $k=3$.
If we replace $F(C)$ with (\ref{F(C)Gr}) according our previous discussion we will obtain
\begin{eqnarray}
A^*_{3,4+1}&=&\int_{\Gamma} d\tau \tilde{F}'(C),~\mbox{where}~\tilde{F}'(C)=\frac{(612)}{(512)}\frac{1}{(123)\ldots(612)}
\end{eqnarray}
which is equivalent to (\ref{ExampleA*NMHVcontour}) as expected, after appropriate choice of $\Gamma$, which should encircle poles at $(123)$, $(345)$ and $(561)$.
Also from this example we see, that in the case of $A^*_{3,n+1}$ off-shell amplitudes we can explicitly construct integration contours for their Grassmannian integral representations (i.e. the maps $\textbf{S}=(\tilde{S}_{6},\ldots,\tilde{S}_{n+2})$, $\textbf{S}:\mathbb{C}^{(n-3)} \mapsto \mathbb{C}^{(n-3)}$, whose zeros determine the integration contours $\Gamma_{3,n+2}^{tree}$ in (\ref{offshellAmpGrassmannian})). The latter are given by:
\begin{eqnarray}
\tilde{S}_j&=&(j-2j-1j)(j12)(23j-2),~j=6,\ldots,n+1,\nonumber\\
\tilde{S}_{n+2}&=&(nn+1n+2)(n+112)(23n).
\end{eqnarray}
This expression is easily obtained by considering integration contour $(\tilde{S}_6,\ldots,\tilde{S}_{n+2})$ for $n+2$ point on-shell amplitude and accounting for $Reg.\sim (n+212)/(n+112)$ factor. It is believed, that in the case
$k >3$ the integration contours can be constructed in similar fashion.

In the end of this section let's make the following remark. Conditions (\ref{S(t)conic})
in general and for $n+2=6$ in particular can be interpreted as conditions for
6 points lie on a conic. For example:
\begin{eqnarray}
S(1)=(123)(345)(561)(246)-(234)(456)(612)(351)=0
\end{eqnarray}
is condition that 6 points in Grasmannian $\mathbb{C}\mathbb{P}^2$ lie on a single conic (any general 5 points determines a conic, so this is condition that point 6 also belongs to the conic). It is natural to ask if there is any geometrical interpretation
for
\begin{eqnarray}
S&=&(123)(345)(561)(245)-(234)(456)(512)(351)=0
\end{eqnarray}
which appears in our construction. We found that conditions on matrix $C$ to be of Veronese form
and $S=0$ are equivalent to $\sigma_6=\sigma_5$ and $\sigma_1,\ldots,\sigma_5$ are arbitrary (here as in \cite{UnificationResidues} we rescaled all $s_i$ to 1) i.e. we interpret this condition
as point 6 belongs to the conic and coincides with point 5. This is probably not very surprising. The dimensionality of the Grassmannian is related to the number of independent kinematical variables.
In the case of off-shell amplitudes (Wilson line form factors) we use axillary spinors
$\lambda_p$ and $\lambda_{\xi}$ in description of off shell momenta. But in the final result dependence on $\lambda_{\xi}$ drops out \cite{offshell-1leg,offshell-multiplelegs,vanHamerenBCFW1}, so effectively we have less variables than naively expected.

\section{Gluing procedure and amplitudes}\label{GluingAmplitudesSec}
Let's return once again to the formula (\ref{offshellamp2}). We have seen in the previous sections that if we leave integrals with respect to $\beta_{1,2}$ intact and concentrate on integrations with respect to $\d s_a\d\sigma_a$ the result (after appropriate choice of integration contours) will be proportional to $A_{k,n+2}$ on-shell amplitude where the dependance on $\beta_{1,2}$ is condensed in kinematical variables
$\{\lambda_i,\tilde{\lambda}_i,\eta_i\}_{i=n+1}^{n+2}$ (see (\ref{spinorsAsFunctionsOfBeta}) and (\ref{kDecompsition})).

So we can think of some integral operator $\hat{A}$ which directly transforms on-shell amplitudes into
Wilson line form factors and correlation functions:
\begin{eqnarray}
\hat{A}:	A_{k,n+2} \mapsto A^*_{k,n+1}.
\end{eqnarray}
We will call this operator \emph{the gluing operator} and will label it $\hat{A}_{i,i+1}$. Label $i$ corresponds to the position of kinematical variables on which this operator acts. Another way to introduce this operator is simply consider convolution (in the seance of discussion in the beginning of section \ref{StringVertexesCorrelationFunctionsSec}) of $A^*_{2,2+1}$ minimal off-shell amplitude with some function of $\{\lambda_i,\tilde{\lambda}_i,\eta_i\}$. In this sense
\emph{the gluing operation} (for Wilson line form factors and correlation functions) discussed in section \ref{StringVertexesCorrelationFunctionsSec} is given by the action of \emph{gluing operator}.

From practical point of view it is useful
because one can immediately utilise large library of answers for on-shell amplitudes into
Wilson line form factors and correlation functions, which in turn can be interpreted as Reggeon
amplitudes.

Note also that similar procedure should work for the form factors of operators from
stress tensor supermultiplet, i.e. for different choice of $\psi$ "target space wave function"
which participates in gluing procedure. Though we will not discuss it in details, and leave this
topic or separate publication.
Here we will concentrate on the simplest case of Wilson line form factors and correlation functions.

So more formally let's \emph{define} gluing operator $\hat{A}_{n+1,n+2}[...]$ acting on the space of functions $f$ of $\{\lambda_i,\tilde{\lambda}_i,\eta_i\}_{i=1}^{n+2}$ variables as
\begin{eqnarray}
\hat{A}_{n+1,n+2}[f]\equiv\int\prod_{i=n+1}^{n+2} \frac{d^2\lambda_{i}d^2\tilde{\lambda}_{i}d^4\eta}{\mbox{Vol}[GL(1)]} ~A^*_{2,2+1}~\times~f\left(\{\lambda_i,\tilde{\lambda}_i,\eta_i\}_{i=1}^{n+2}\right).
\end{eqnarray}
Performing integration over  $\vlt_{n+1}$, $\vlt_{n+2}$, $\vlet_{n+1}$ and $\vlet_{n+2}$ variables as in Appendix \ref{appA} we get
\begin{eqnarray}
\hat{A}_{n+1,n+2}[f]=\frac{\langle p \xi\rangle}{\kappa^*}\int \frac{d\beta_1}{\beta_1}
\wedge\frac{d\beta_2}{\beta_2}~\frac{1}{\beta_1^2\beta_2}  ~ f\left(\{\lambda_i,\tilde{\lambda}_i,\tilde{\eta}_i\}_{i=1}^{n+2}\right)\big{|}_{*},
\end{eqnarray}
where $\big{|}_{*}$ denotes substitutions $\{\lambda_i,\tilde{\lambda}_i,\eta_i\}_{i=n+1}^{n+2}
\mapsto\{\lambda_i(\beta),\tilde{\lambda}_i(\beta),\tilde{\eta}_i(\beta)\}_{i=n+1}^{n+2}$ with
\begin{align}
&\vll_{n+1}(\beta) = \vlluu_{n+1} + \beta_2\vlluu_{n+2}\, , && \vlt_{n+1}(\beta) =
\beta_1\vltuu_{n+1}  + \frac{(1+\beta_1)}{\beta_2}\vltuu_{n+2}\, ,
&&\vlet_{n+1}(\beta) = -\beta_1\vleuu_{n+1}\, , \nonumber\\
&\vll_{n+2}(\beta) = \vlluu_{n+2} + \frac{(1+\beta_1)}{\beta_1\beta_2}\vlluu_{n+1}\, ,
&& \vlt_{n+2}(\beta) = -\beta_1\vltuu_{n+2} -\beta_1\beta_2 \vltuu_{n+1}\, , &&\vlet_{n+2}(\beta) = \beta_1\beta_2\vleuu_{n+1}\, .
\end{align}
and
\begin{eqnarray}
\vlluu_{n+1}=\lambda_p,~\vltuu_{n+1}=\frac{\la\xi |k}{\la\xi p\ra},~\vleuu_{n}=\vlet_p;
~~\vlluu_{n+2}=\lambda_{\xi},~\vltuu_{n+2}=\frac{\la p |k}{\la\xi p\ra},~\vleuu_{n+2}=0.
\end{eqnarray}
Here we also understand integration with respect to $\beta_{1,2}$ as residue form, and
will always evaluate it at points $res_{\beta_2=0}\circ
res_{\beta_1=-1}$.
After this formal introduction we are ready to consider several examples of action of $\hat{A}_{i,i+1}$ on on-shell amplitudes.

\subsection{Tree level}\label{GluingAmplitudesSec1}
So let's test how our gluing operator works on some explicit examples. The simplest case is given by the action of $\hat{A}$ on  $k=2$, $n+2=4$ point amplitude $A_{2,4}$.
We expect that we will reproduce $A_{2,2+1}^*$ off-shell amplitude (Wilson line operator form factor) by the action of operator $\hat{A}_{34}[\ldots]$ on $A_{2,4}$ on-shell amplitude. Note also, that  next steps are actually identical for all $k=2$ amplitudes with arbitrary $n$.
Indeed in the case of $A_{2,4}$ amplitude we have
\begin{eqnarray}
A_{2,4}(\Omega_1,\ldots,\Omega_4)=\delta^4(p_{1234})\frac{\delta^8(q_{1234})}{\langle12\rangle\langle23\rangle\langle34\rangle\langle41\rangle}.
\end{eqnarray}
Introducing notations\footnote{Here for simplicity we also drop spinorial and $SU(4)_R$ indices.}
\begin{eqnarray}
p_{1\ldots n}\equiv\sum_{i=1}^np_i\equiv\sum_{i=1}^n\lambda_i\tilde{\lambda}_i,~p_{1\ldots n}^2=p_{1,n}^2,~
q_{1\ldots n}\equiv\sum_{i=1}^n\lambda_i\tilde{\eta}_i,
\end{eqnarray}
the $A_{2,4}$ amplitude with $|_{*}$ substitution applied takes the form
\begin{eqnarray}
A_{2,4}\Big{|}_{*}=\delta^4(p_{12}+k)\frac{\delta^8(q_{12p})\beta_1^2\beta_2}
{\langle12\rangle(\langle2p\rangle+\beta_2\langle2\xi\rangle)\langle
	p\xi\rangle
	(\beta_1\beta_2\langle1\xi\rangle+(1+\beta_1)\langle1p\rangle)}.
\end{eqnarray}
Now, evaluating integral over $\beta_1,\beta_2$ by means of composite residue
$res_{\beta_1=-1}\circ res_{\beta_2=0}[...]$ we get
\begin{eqnarray}\label{MHVGluingExample}
\hat{A}_{34}[A_{2,4}(\Omega_1,\ldots,\Omega_4)]=\delta^4(p_{12}+k)
\frac{\delta^8(q_{12p})}{\kappa^*\langle12\rangle}~res_{\beta_2=0}\circ
res_{\beta_1=-1}[\omega]=A^*_{2,2+1}(\Omega_1,\Omega_2,g_3^*),\nonumber\\
\end{eqnarray}
where
\begin{eqnarray}
\omega=\frac{d\beta_2\wedge d\beta_1}{\beta_2\beta_1(\langle2p\rangle+\beta_2\langle2\xi\rangle)(\beta_1\beta_2\langle1\xi\rangle+(1+\beta_1)\langle1p\rangle)},
\end{eqnarray}
and (the projector $\partial_{\eta_p}^4$ acting on $A^*_{2,n+1}$ is implemented)
\begin{eqnarray}
A^*_{2,2+1}(\Omega_1,\Omega_2,g_3^*)=\delta^4(p_{12}+k)\frac{1}{\kappa^*}
\frac{\delta^8(q_{12p})}{\langle p1\rangle\langle 12\rangle\langle 2p\rangle}.
\end{eqnarray}
This is in agreement with results of \cite{offshell-1leg,vanHamerenBCFW1} for $A^*_{2,2+1}$.

Proceeding in similar way let's consider action of $\hat{A}_{n+1n+2}$ on $A_{2,n+2}$ on-shell amplitude. We expect that the result will be
given by $A^*_{2,n+1}$ Wilson line form factor. Indeed it is easy\footnote{It can be obtained by simple spinor relabellings from previous example}  to see that
\begin{eqnarray}
\hat{A}_{n+1n+2}[A_{2,n+2}(\Omega_1,\ldots,\Omega_{n+2})]=A^*_{2,n+1}(\Omega_1,\ldots,\Omega_{n},g_{n+1}^*)=\frac{\delta^4(p_{1\ldots n}+k)}{\kappa^*}
\frac{\delta^8(q_{1\ldots p})}{\langle p1\rangle\langle 12\rangle\ldots\langle np\rangle}.
\nonumber\\
\end{eqnarray}
This is once more in agreement with results of \cite{offshell-1leg,vanHamerenBCFW1} for $A^*_{2,n+1}$.

Proceeding further in similar way we can also reproduce another results obtained from BCFW
recursion \cite{offshell-1leg,vanHamerenBCFW1} for component off-shell amplitudes (Wilson line operator form factors) $A^*_{3,3+1}(1^-2^-3^+g_4^*)$ and $A^*_{3,4+1}(1^+2^+3^-4^-g_5^*)$ \cite{offshell-1leg,vanHamerenBCFW1}. We expect to obtain them from the on shell amplitudes
$A_{3,5}(1^-2^-3^+4^-5^+)$ and $A_{3,6}(1^+2^+3^-4^-5^-6^+)$ by means of action of $\hat{A}_{45}[\ldots]$ and $\hat{A}_{56}[\ldots]$ correspondingly. Performing simple computations
(the explicit details and answers can be found in Appendix \ref{appC}) we see that indeed the following relations holds:
\begin{eqnarray}
\hat{A}_{45}[A_{3,5}(1^-2^-3^+4^-5^+)]&=&A^*_{3,3+1}(1^-2^-3^+g_4^*),\\
\hat{A}_{56}[A_{3,6}(1^+2^+3^-4^-5^-6^+)]&=&A^{*}_{3,4+1}(1^+2^+3^-4^-g_5^*),
\end{eqnarray}
in agreement with previously obtained results for Willson line form factors \cite{offshell-1leg,vanHamerenBCFW1}.

As a final example we would like to consider the case with multiple gluing operations applied.
Let's consider quite  none trivial example of such situation, namely let's consider
correlation function of three Wilson line operators $A^*_{3,0+3}(g_1^*,g_2^*,g_3^*)$.
According our construction of generalised worldsheet vertex operators it should be given by:
 \begin{equation}
A^*_{3,0+3}(g_1^*,g_2^*,g_3^*)=\left\la {\mathcal V}^{\text{WL}}_{1,
	2}{\mathcal V}^{\text{WL}}_{3, 4}{\mathcal V}^{\text{WL}}_{5,
	6}\right\ra\, .
\end{equation}
It was first computed in \cite{vanHamerenBCFW1} by means of BCFW recursion and later reproduced
in \cite{offshell-multiplelegs} form Grassmannian integral representation. The result is given by:
\begin{eqnarray}
A^*_{3,0+3}(g^*_1,g^*_2,g^*_3)&=&\delta^4(k_1+k_2+k_3)
\left(1+\mathbb{P}'+\mathbb{P}'^2\right)\tilde{f},\nonumber\\
\tilde{f}&=&\frac{\langle
	p_1p_2\rangle^3[p_2p_3]^3}{\kappa_3\kappa^*_1\langle p_2|k_1|p_3]\langle
	p_1|k_3|p_2]\langle p_2|k_1|p_2]}.
\end{eqnarray}
Here $\mathbb{P}'$ is permutation operator which now  shifts all spinor and momenta labels by +1 mod 3.

We want to show that $A^*_{3,0+3}$ can be reproduced from $\mbox{NMHV}_6$
point on shell amplitude $A_{3,6}(1^-2^+3^-4^+5^-6^+)$ by application of the following product of gluing operators $\hat{A}_{12}\circ\hat{A}_{34}\circ\hat{A}_{56}$.
Indeed (see appendix \ref{appC} for details), the following relation holds:
\begin{equation}
A^*_{3,0+3}(g_1^*,g_2^*,g_3^*)
=(\hat{A}_{12}\circ\hat{A}_{34}\circ\hat{A}_{56})[A_{3,6}(1^-2^+3^-4^+5^-6^+)]] ,
\end{equation}
where $A_{3,6}(1^-2^+3^-4^+5^-6^+)$ amplitude is given by
\begin{equation}
A_{3,6}=\delta^4(p_{1\ldots6})
\left(1+\mathbb{P}^2+\mathbb{P}^4\right)f,
~f=\frac{\langle13\rangle^4[46]^4}{\langle12\rangle\langle23\rangle[45][56]\langle3|1+2|6]\langle1|5+6|4]p_{456}^2}
\end{equation}
and $\mathbb{P}$ is permutation operator shifting spinor labels by +1 mod 6.

So we have seen on multiple examples that presented in the beginning of this section
gluing operator $\hat{A}_{i,i+1}$ allows one to convert on-shell amplitudes into Willson line
form factors and correlation functions at tree level.

At the end of this subsection we would like to point out the formal analogy between the action of $\hat{A}_{ii+1}$ operators on $A_{k,n}$ amplitudes and action of $R$-matrices on some vacuum state of integrable spin chain. Indeed,  it looks like $\hat{A}_{ii+1}$ operator creates "excitation" (Wilson-line operator insertion) in the "vacuum" consisting from on-shell states. We think that this analogy being properly investigated may provide us with the answer to question "what is the appropriate description of Wilson line form factors in terms of some integrable system?".

It is also interesting to note that the integration with respect to $\beta_1,\beta_2$ variables, which was performed by taking residues, is in fact equivalent to the choice of specific kinematical limit for the momenta
$p_{n+1}$ and $p_{n+2}$ of the initial on-shell amplitude. If we naively take consecutive
limits $\beta_1 \rightarrow -1,\beta_2 \rightarrow 0$ in the definitions of momenta  $p_{n+1}(\beta)$ and $p_{n+2}(\beta)$, which is equivalent to residue evaluation, we would get finite result\footnote{Note, that if we would take the limits in opposite order the results for $p_{n+1}$, $p_{n+2}$ momenta would diverge, but the expression for off-shell amplitudes would still be finite.}
\begin{eqnarray}
p_{n+1}&=\vlluu_{n+1}\vltuu_{n+1},~p_{n+2}=\vlluu_{n+2}\vltuu_{n+2}.
\end{eqnarray}
On the other hand, for all $\omega$ forms which we have encountered in previous examples we may use global residue theorem to relate multiple residue at $\beta_1 = -1, \beta_2 = 0$  with the multiple residue at  $\beta_1 = 0, \beta_2 = 0$. If we take the limits $\beta_1 \rightarrow 0,\beta_2 \rightarrow 0$ (regardless of the order of limits) in the definitions of $p_{n+1}(\beta)$ and $p_{n+2}(\beta)$ momenta we will get singular result
\begin{eqnarray}
p_{n+1}&=&
\frac{1}{\beta_2}\vlluu_{n+1}\vltuu_{n+2}
+\vlluu_{n++2}\vltuu_{n+2}+O(\beta_2),\nonumber\\
p_{n+2}&=&\frac{1}{\beta_2}\vlluu_{n+1}\vltuu_{n+2}
+\vlluu_{n+1}\vltuu_{n+1}+O(\beta_2),
\end{eqnarray}
which is equivalent to BCFW shift $[n+1,n+2\rangle$ of $p_{n+1}=\vlluu_{n+2}\vltuu_{n+2}$ and
$p_{n+2}=\vlluu_{n+1}\vltuu_{n+1}$ momenta evaluated at large $z$. The behavior of amplitudes in the limit $z \rightarrow \infty$ may be interpreted as special kinematical limit with some particles with large (complex) light like momenta traveling in the soft background\cite{ArkaniHamedOnTreeAmplitudesinGaugeTheoryandGravity}. So in this sense our
gluing procedure is closely related to the specific high energy kinematical limit of the ordinary on-shell amplitudes.

\subsection{Integrands}\label{GluingAmplitudesSec2}
So far we have seen that by means of gluing operator $\hat{A}_{i,i+1}$ we can to convert on-shell amplitudes into Willson line form factors and correlation functions at tree level, formally without any reference to ambitwistor string or Grassmannian representation. Though the explicit form of this operator is of course motivated heavily by our construction of generalised vertex operators.

But what about loops? Since our
gluing operator acts in simple manner on rational functions it is natural to try verify conjecture that using $\hat{A}_{i,i+1}$ we can convert integrands of on-shell amplitudes into integrands of
Wilson line form factors and correlation functions as  well.

To see whether this conjecture is reasonable let us consider the simplest possible example of $k=2$, $n+1=3$ point one loop amplitude $A^{*(1)}_{2,2+1}$ and show that the gluing operator applied to the integrand of $A^{(1)}_{2,4}$ amplitude will give us the desired expression for the integrand of $A^{*(1)}_{2,2+1}$ amplitude.

But first we need to obtain the integrand of
$A^{*(1)}_{2,2+1}$ amplitude itself. The easiest way to get it is try to reconstruct it
from $k^2$ channel unitarity cut. Considering the latter (taking residues of the
integrand with respect to the poles of $1/l^2_1$ and
$1/l^2_2$ propagators, see Fig. \ref{fig:integrand1}) we have\footnote{The necessary manipulations are similar to the case of $s$-channel cut of $A_{2,4}^{(1)}$ amplitude. $(pq)$
here stands for standard Mincovskian scalar product.}:
\begin{eqnarray}
A^{*(1)}_{2,2+1}\Big{|}_{k^2~cut}=\int d^4\eta_{l_1}d^4\eta_{l_2}
A^{*}_{2,2+1}(l_1,l_2,g^*)A_{2,4}(l_1,l_2,\Omega_2,\Omega_1)=A^{*}_{2,2+1}(l_1,l_2,g^*)
\frac{Tr(kp21)}{(pl_2)(l_22)}.\nonumber\\
\end{eqnarray}
The $Tr$ factor can be transformed into
$k^2(p+p_2)^2=(p_1+p_2)^2(p_2+p)^2$ with the help of momentum
conservation $k+p_1+p_2=0$, and $k_T$ decomposition conditions $(pk)=0$. Thus, the expression for $A^{*(1)}_{2,2+1}(\Omega_1,\Omega_2,g^*)$ amplitude is given by
\begin{eqnarray}
A^{*(1)}_{2,2+1}(\Omega_1,\Omega_2,g^*)=A^{*}_{2,2+1}(\Omega_1,\Omega_2,g^*)\int d^Dl
\frac{(p_1+p_2)^2(p_2+p)^2}{l^2(l+p_2)^2(l+p_1+p_2)^2(pl)},
\end{eqnarray}
which contains one loop scalar box integral with one of the propagators, namely $1/l^2$ replaced by its eikonal counterpart $1/(pl)$, see Fig. \ref{fig:integrand2}.
\begin{figure}[h]
	\centering
	\begin{gather}
	\begin{tikzpicture}[baseline={($(c1.base) - (0,0.1)$)},transform shape, scale=0.8]
	\coordinate (c1) at (0,0);
	\coordinate (c2) at (2,0);
	\coordinate (c3) at (2,0.9);
	\coordinate (c4) at (2,-0.9);
	\coordinate (c5) at (5,0.9);
	\coordinate (c6) at (5,-0.9);
	\node[right] (l1) at (3.5,1.3) {\large $l_1$};
	\node[right] (l2) at (3.5,-1.4) {\large $l_2$};
	\node[above] (p1) at (7.5,2) {\large $p_1$};
	\node[below] (p2) at (7.5,-2) {\large $p_2$};
	\node[above] (nk) at (0,0.2) {\large $k$};
	\draw[dashed,very thick] (3.5,-1.7) -- (3.5,1.7);
	\draw[very thick,->-] (c5) -- (p1);
	\draw[very thick,->-] (c6) -- (p2);
	\draw[very thick] (c3) -- (c5);
	\draw[very thick] (c4) -- (c6);
	\draw[ultra thick,->-] (1.5,0) -- (-1,0);
	\fill[black] (-1,-0.05) rectangle (2,0.05);	
	\node[ellipse, black, fill=grayn, minimum width=1.5 cm, minimum height=3 cm, draw, inner sep=0pt] at (2,0) {};
	\node[ellipse, black, fill=grayn, minimum width=1.5 cm, minimum height=3 cm, draw, inner sep=0pt] at (5,0) {};
	\end{tikzpicture}
	\nonumber
	\end{gather}
	\caption{Unitarity cut of $A^{*(1)}_{2,2+1}$ amplitude in $k^2=(p_1+p_2)^2$ channel.}
	\label{fig:integrand1}
\end{figure}
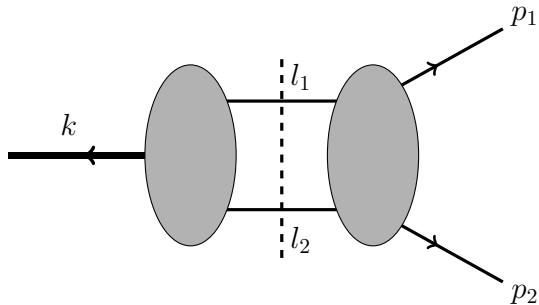
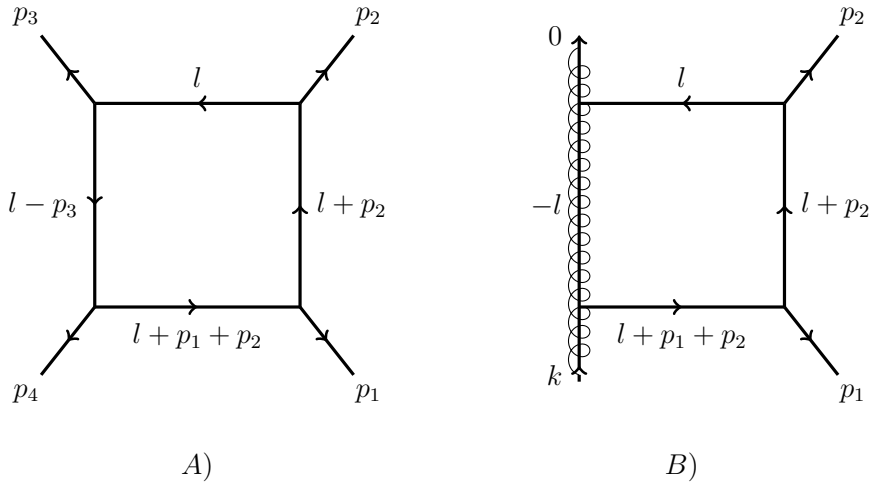
\begin{figure}[h]
	\centering
	\begin{gather}
	\begin{tikzpicture}[baseline={($(n1.base) - (0,0.1)$)},transform shape, scale=0.9]
	\coordinate (n4) at (0,0);
	\coordinate (n1) at (3,0);
	\coordinate (n2) at (3,3);
	\coordinate (n3) at (0,3);
	\node[below] (p1) at (4,-1) {$p_1$};
	\node[above] (p2) at (4,4) {$p_2$};
	\node[above] (p3) at (-1,4) {$p_3$};
	\node[below] (p4) at (-1,-1) {$p_4$};
	\node[above] at (1.5,3.1) {$l$};
	\node[right] at (3.1,1.5) {$l+p_2$};
	\node[left] at (-0.1,1.5) {$l-p_3$};
	\node[below] at (1.5,-0.1) {$l+p_1+p_2$};
	\node[below] at (1.5,-2) {$A)$};
	\draw[very thick,->-] (n4) -- (n1);
	\draw[very thick,->-] (n1) -- (n2);
	\draw[very thick,->-] (n2) -- (n3);
	\draw[very thick,->-] (n3) -- (n4);
	\draw[very thick,->-] (n1) -- (p1);
	\draw[very thick,->-] (n2) -- (p2);
	\draw[very thick,->-] (n3) -- (p3);
	\draw[very thick,->-] (n4) -- (p4);
	\end{tikzpicture}
	\qquad\qquad
	\begin{tikzpicture}[baseline={($(n1.base) - (0,0.1)$)},transform shape, scale=0.9]
	\coordinate (n4) at (0,0);
	\coordinate (n1) at (3,0);
	\coordinate (n2) at (3,3);
	\coordinate (n3) at (0,3);
	\coordinate (c0) at (0,4);
	\coordinate (ck) at (0,-1);
	\coordinate (ckm) at (0,-1.1);
	\node[left] (n0) at (-0.1,4) {$0$};
	\node[left] (nk) at (-0.1,-1) {$k$};
	\node[below] (p1) at (4,-1) {$p_1$};
	\node[above] (p2) at (4,4) {$p_2$};
	\node[above] at (1.5,3.1) {$l$};
	\node[right] at (3.1,1.5) {$l+p_2$};
	\node[left] at (-0.1,1.5) {$-l$};
	\node[below] at (1.5,-0.1) {$l+p_1+p_2$};
	\node[below] at (1.5,-2) {$B)$};
	\draw[very thick,->-] (n4) -- (n1);
	\draw[very thick,->-] (n1) -- (n2);
	\draw[very thick,->-] (n2) -- (n3);
    \draw[very thick] (n4) -- (n3);
    \draw[very thick,->] (n3) -- (c0);
	\draw[very thick,->-] (n1) -- (p1);
	\draw[very thick,->-] (n2) -- (p2);
	\draw[very thick,>-] (ck) -- (n4);
	\draw[very thick] (ckm) -- (ck);
	\draw[gluon] (ck) -- (c0);
	\end{tikzpicture}
	\nonumber
	\end{gather}
	\caption{Scalar box integrals contributing to $A^{(1)}_{2,2+1}$ and $A^{*(1)}_{2,2+1}$ amplitudes correspondingly. Line with coil denotes eikonal propagator $1/(pl)$.}
	\label{fig:integrand2}
\end{figure}

Now let's turn to the integrand of  $A^{(1)}_{2,4}$ amplitude:
\begin{eqnarray}
A^{(1)}_{2,4}(\Omega_1,\ldots,\Omega_4)=A_{2,4}(\Omega_1,\ldots,\Omega_4)\int d^Dl
\frac{(p_1+p_2)^2(p_2+p_3)^2}{l^2(l+p_2)^2(l+p_1+p_2)^2(l-p_3)^2}.
\end{eqnarray}
It should be noted, that the notion of integrand is uniquely defined only in dual variables. So, to be accurate we should consider the gluing operation in such variables (momentum twistors) also. Here, we will however use helicity spinors in a hope that possible loop momentum rearrangement will not cause any trouble. It turns out that it is indeed the case as we will see in a moment. Acting with $\hat{A}_{34}$ operator on the $A^{(1)}_{2,4}(1,2,3,4)$ integrand
\begin{eqnarray}
Int=A_{2,4}(\Omega_1,\ldots,\Omega_4)
\frac{(p_1+p_2)^2}{l^2(l+p_2)^2(l+p_1+p_2)^2}\frac{(p_2+p_3)^2}{(l-p_3)^2}.
\end{eqnarray}
and using momentum definitions (\ref{momentaAsFunctiosFoBeta}) we get
\begin{eqnarray}
\hat{A}_{34}[Int]=A^{*}_{2,2+1}(\Omega_1,\Omega_2,g_3^*)
\frac{(p_1+p_2)^2}{l^2(l+p_2)^2(l+p_1+p_2)^2}\frac{(p_2+p)^2}{(lp)},
\end{eqnarray}
which is exactly the integrand expression for  $A^{*(1)}_{2,2+1}$ amplitude.

This example gives us a hope that more accurate and general consideration of gluing procedure at the level of integrands will be also successful and will provide us with the prescription for obtaining
$A^{*(l)}_{k,m+n}$ integrands from the corresponding $A^{(l)}_{k,m+2n}$ integrands by application of appropriate combinations of
$\hat{A}_{ii+1}$ operators.

\section{Conclusion}\label{ConclusionSec}

In this paper we presented results for derivation of scattering equations (RSV) and Grassmannian representations  for form factors of local and Wilson line operators in $\mathcal{N}=4$ SYM from corresponding four dimensional ambitwistor string theory. In the case of local operators we restricted ourselves to the case of form factors  of operators from stress-tensor operator supermultiplet. The obtained results are in agreement with previously obtained Grassmannian integral representations. As by product we discovered an easy and convenient gluing procedure, which allows to obtain required form factor expressions from already known amplitude expressions. The construction of composite string vertex operators for the analysed local or Wilson line operators was inspired by the mentioned gluing procedure.  An interesting future research direction, which we are planning to pursue, will be to consider pullbacks of composite operators defined on twistor or Lorentz harmonic chiral superspace \cite{TwistorFormFactors1,TwistorFormFactors2,TwistorFormFactors3,LHC1,LHC2,LHC3,LHC4}. We hope that our consideration along these lines could be extended to arbitrary local composite operators.

Next, it would be very interesting to fully uncover geometrical picture behind Grassmannian and scattering equations representations for form factors of local and Wilson line operators. It is interesting to see if the ``Amplituhedron'' picture could be extended to all possible gauge invariant observables in  $\mathcal{N}=4$ SYM, which representations under global gauge transformations may differ from singlet representation.

Having obtained scattering equations representations one may wonder what is the most efficient way to
get final expressions for particular form factors with given number of particles and their helicities. In the case of amplitudes we know that it is given by computation of global residues with the methods of computational algebraic geometry \cite{GlobalResidue1,GlobalResidue2,GlobalResidue3} , see also \cite{Talesof1001Gluons}.  It would be interesting to see how this procedure works in the case of tree level form factors, their loop level integrands and provide necessary details needed when writing computer codes.

Finally, it would be interesting to consider loop corrections to form factors of Wilson line operators (gauge invariant off-shell amplitudes). Also, it is extremely interesting to see how the presented approach works in other theories, for example in gravity and supergravity, where in the case of gauge invariant off-shell amplitudes we have a well developed framework based on high-energy effective lagrangian \cite{gravityEL1,gravityEL2,gravityEL3,gravityEL4,gravityEL5}, see also \cite{gravityRegge1,gravityRegge2,gravityRegge3,gravityRegge4,gravityRegge5} for similar research along this direction.

\section*{Acknowledgements}

The authors would like to thank D.I. Kazakov, L.N. Lipatov  and  Yu-tin Huang for interesting and stimulating discussions. This work was supported by RFBR grants \# 17-02-00872, \# 16-02-00943 and contract \# 02.A03.21.0003 from 27.08.2013 with Russian Ministry of Science and Education.

\appendix

\section{Gluing procedure and Grassmannians}\label{appA}
In this appendix we are going to present computational details from section 4
of the main text.

Before proceeding with actual formulas let's make the following remark.
The gluing procedure (gluing operator) introduced in \cite{FormFactorsGrassmanians} was in fact implicitly used to obtain both the Grassmannian integral representation for form factors of operators from stress-tensor supermultiplet \cite{FormFactorsGrassmanians} and off-shell  amplitudes with one leg off-shell \cite{offshell-1leg}. The idea was to take a top-cell diagram for amplitude, perform a sequence of square and merge/unmerge moves until we get a box (assume it is possible for the diagram under consideration) on the boundary and replace it with the corresponding minimal form factor or off-shell amplitude. Graphically, this relation reads\footnote{We have borrowed this nice picture from \cite{FormFactorsGrassmanians}}
\begin{equation}
\scalebox{0.9}{\(
	\begin{aligned}
	\begin{tikzpicture}[scale=0.8, baseline=-0.7cm]
	\draw (1,1) -- (1,2) -- (2,2) -- (2,1) -- (1,1);
	\draw (1,0) -- (1,1);
	\draw (2,0) -- (2,1);
	\draw (2.5,2.5) -- (2,2);
	\draw (0.5,2.5) -- (1,2);
	\draw (-0.5,-0.75) -- (-0.5,-2.5);
	\draw (1.5,-0.75) -- (1.5,-2.5);
	\draw (2.5,-0.75) -- (2.5,-2.5);
	\draw (3.5,-0.75) -- (3.5,-2.5);
	\node at (-0.5,-2.5-\labelvdist) {$n$};
	\node at (0.5,-2.5-\labelvdist) {$\cdots$};
	\node at (1.5,-2.5-\labelvdist) {$3$};
	\node at (2.5,-2.5-\labelvdist) {$2$};
	\node at (3.5,-2.5-\labelvdist) {$1$};
	\node at (2.5+\labelddist,2.5+\labelddist) {$n+2$};
	\node at (0.5+\labelddist,2.5+\labelddist) {$n+1$};
	\node[dw] at (2,1) {};
	\node[dw] at (1,2) {};
	\node[db] at (1,1) {};
	\node[db] at (2,2) {};
	\node[ellipse, black, fill=grayn, minimum width=4 cm, minimum height=2 cm, draw, inner sep=0pt] at (1.5,-0.75) {};
	\end{tikzpicture}
	\quad
	\longrightarrow
	\quad
	\begin{tikzpicture}[scale=0.8, baseline=-0.7cm]
	\draw[thick,double] (1.5,-0+1.5) -- (1.5,-0.5+1.5); 
	\draw (1.5,-0.5+1.5) -- (2,-\vacuumheight+1.4);
	\draw (1.5,-0.5+1.5) -- (1,-\vacuumheight+1.4);
	\draw (-0.5,-0.75) -- (-0.5,-2.5);
	\draw (1.5,-0.75) -- (1.5,-2.5);
	\draw (2.5,-0.75) -- (2.5,-2.5);
	\draw (3.5,-0.75) -- (3.5,-2.5);
	\node at (-0.5,-2.5-\labelvdist) {$n$};
	\node at (0.5,-2.5-\labelvdist) {$\cdots$};
	\node at (1.5,-2.5-\labelvdist) {$3$};
	\node at (2.5,-2.5-\labelvdist) {$2$};
	\node at (3.5,-2.5-\labelvdist) {$1$};
	\node[ellipse, black, fill=grayn, minimum width=4 cm, minimum height=2 cm, draw, inner sep=0pt] at (1.5,-0.75) {};
	\fill[black!60!white] (1.43,1.95) rectangle (1.56,0.9);
	\end{tikzpicture}
	\quad
	\eqncom
	\end{aligned}
	\)} \label{amplitudeToformfactor}
\end{equation}
where the box at the legs $n+1$ and $n+2$ was replaced for the sake of concreteness. We got a similar relation of form factor on-shell diagrams to the amplitude on-shell diagrams in \cite{SoftTheoremsFormFactors,q2zeroFormFactors} based on soft limit procedure. The corresponding box diagram was deformed by extra soft factor, so that it became equivalent to the corresponding minimal form factor.

It turns out however, that there is a simpler gluing procedure, which we used in our consideration. Namely, we can glue (perform the on-shell phase space integration - perform convolution with) minimal form factor or off-shell amplitude directly to the amplitude top cell diagram without cutting off mentioned above boxes.
%

Now let's return to actual computations.
Let's consider the case of Wilson line operators first. Let's consider (\ref{offshellamp1}) once more.
If we choose $F(C)$ according to (\ref{F(C)GR}) we can rewrite (\ref{offshellamp1}) as:\footnote{Without loss of generality we may choose off-shell leg to lie between legs $1$ and $n$.}
\begin{eqnarray}
A^{*}_{k,n+1} = \int\prod_{i=n+1}^{n+2}\frac{d^2\vll_i d^2\vlt_i}{\text{Vol}[GL(1)]} d^4\vlet_i A_{2,2+1}^*(g^{*},\Omega_{n+1},\Omega_{n+2})\Big|_{\vll\to -\vll} A_{k,n+2}\, ,
\end{eqnarray}
where $A_{k,n+2}$ stands for corresponding Grassmannian integral - top-cell diagram (which can be evaluated into N$^{k-2}$MHV $n+2$ point on-shell amplitude if the appropriate integration contour is chosen)\footnote{Here we left the integration contour unspecified.}:
\begin{eqnarray}
A_{k,n+2} = \int \frac{d^{k\times (n+2)} C}{\text{Vol}[GL(k)]}
\frac{\delta^{k\times 2} (C\cdot\vlt)\delta^{k\times 4} (C\cdot\vlet)\delta^{(n+2-k)\times 2} (C^{\perp}\cdot\vll)}{(1\cdots k)(2\cdots k+1)\cdots (n+2\cdots k-1)} , \label{GrassmannianIntegralAmplitudes}
\end{eqnarray}
and the minimal off-shell vertex $A_{2,2+1}^*(g^{*},n+1,n+2)$ is given by \cite{offshell-1leg}:
\begin{gather}
A_{2,2+1}^*(g^*,\Omega_{n+1},\Omega_{n+2}) = \frac{1}{\kappa^{*}}\int
\frac{d^{2\times 3} C}{\text{Vol} [GL(2)]}
\frac{\delta^4 (C\cdot \vltu)\delta^8 (C\cdot \vlet) \delta^4 (C^{\perp}\cdot \vll )}{(p\, n+1)(n+1\, n+2)(n+2\, p)} \nonumber \\ =
 \frac{1}{\kappa^*}\int\frac{\d\beta_1}{\beta_1}\frac{\d\beta_2}{\beta_2}
\delta^2 (\vll_p + \beta_1\vll_{n+1}-\beta_1\beta_2\vll_{n+2})
\delta^2 (\vltu_{n+1}) \delta^2 (\vltu_{n+2})
\nonumber \\ \times
\delta^4 (\vlet_{n+1}+\beta_1\vlet_p)\delta^4 (\vlet_{n+2}+\beta_2\vlet_{n+1})
\end{gather}
with $\vltu_{n+1} = \vlt_{n+1} + \frac{\la n+2|k}{\la n+2 | n+1\ra}$
and $\vltu_{n+2} = \vlt_{n+2} + \frac{\la n+1|k}{\la n+1 | n+2\ra}$. Here $p$ is the off-shell gluon direction and $k$ is its momentum. Now, the integration steps up to final integrations in $\beta_1$ and $\beta_2$ follow closely those in \cite{offshell-1leg}. That is, performing integrations over $\vlt_{n+1}$, $\vlt_{n+2}$, $\vlet_{n+1}$ and $\vlet_{n+2}$ we get
\begin{align}
\vlt_{n+1} &= -\frac{\la n+2|k}{\la n+2|n+1\ra} ,\quad
&\vlt_{n+2} &= -\frac{\la n+1|k}{\la n+1|n+2\ra} , \\ \quad  \vlet_{n+1} &= -\beta_1\vlet_p , \quad &\vlet_{n+2} &= \beta_1\beta_2\vlet_p.
\end{align}
The $\text{Vol} [GL(1)]^2$ redundancy in the remaining integrations over  $\vll$ is removed using their parametrization as in \cite{FormFactorsGrassmanians}:
\begin{equation}
\vll_{n+1}=\refspina-\beta_3 \refspinb
\eqncom\qquad
\vll_{n+2}=\refspinb-\beta_4 \refspina ,
\end{equation}
where $\refspina$ and $\refspinb$ are two arbitrary but linearly independent reference spinors. Then $\abr{n \splus 1 \ssep n \splus 2 }=(\beta_3\beta_4-1)\abr{\refspinbl\refspinal}$,
\begin{eqnarray}
\int\frac{d^2\vll_{n+1}}{\text{Vol}[GL(1)]}\frac{d^2\vll_{n+2}}{\text{Vol}[GL(1)]} = - \la\refspina\refspinb\ra^2 \int d\beta_3 d\beta_4
\end{eqnarray}
and
\begin{align}
A^{*}_{k,n+1} =& \frac{1}{\kappa^{*}}\la\refspina\refspinb\ra^2
\int\frac{d^{k\times (n+2)} C}{\text{Vol}[GL(k)]}
\frac{d\beta_1}{\beta_1}\frac{d\beta_2}{\beta_2}
\frac{d\beta_3 d\beta_4}{(1-\beta_3\beta_4)^2} \nonumber \\
&\times \delta^2 (\lambda_p + \beta_1 (1+\beta_2\beta_4)\refspina
- \beta_1 (\beta_2 + \beta_3)\refspinb ) \nonumber \\
&\times\frac{1}{(1\cdots k)\cdots (n+2\cdots k-1)}\delta^{k\times 2} \left( C'\cdot \vltuu \right)
\delta^{k\times 4} \left(C'\cdot \vleuu \right)
\delta^{(n+2-k)\times 2} \left(C'^{\perp}\cdot \vlluu \right). \label{GrassmannianBeforeb3b4Integration}
\end{align}
Here, we introduced the following notation
\begin{align}
C'_{n+1} &= \frac{1}{1-\beta_3\beta_4}C_{n+1} + \frac{\beta_3}{1-\beta_3\beta_4}C_{n+2},
&C'^{\perp}_{n+1} &= C^{\perp}_{n+1} - \beta_4 C^{\perp}_{n+2}, \nonumber \\
C'_{n+2} &= \frac{1}{1-\beta_3\beta_4}C_{n+2} + \frac{\beta_4}{1-\beta_3\beta_4} C_{n+1},
&C'^{\perp}_{n+2} &= C^{\perp}_{n+2} - \beta_3 C^{\perp}_{n+1},
\end{align}
and
\begin{align}
&\vlluu_i = \vll_i, & i = 1,\ldots  n& , &\vlluu_{n+1} &= \refspina ,
&\vlluu_{n+2} &= \refspinb \nonumber \\
&\vltuu_i = \vlt_i, & i = 1,\ldots  n& ,
&\vltuu_{n+1} &= \frac{\la\refspinb |k}{\la\refspinb\refspina\ra},
&\vltuu_{n+2} &= - \frac{\la\refspina |k}{\la\refspinb\refspina\ra} , \nonumber \\
&\vleuu_i = \vlet_i,  & i = 1,\ldots  n& , &\vleuu_{n+1} &= \vlet_p ,
&\vleuu_{n+2} &= 0 . \nonumber \\
\end{align}
The factor of $1/(1-\beta_3\beta_4)^2$ in  (\ref{GrassmannianBeforeb3b4Integration}) is the Jacobian from reorganizing the $C^\perp\cdot\vll$ delta functions (see \cite{FormFactorsGrassmanians} for further details). Next, rewriting the first delta function in (\ref{GrassmannianBeforeb3b4Integration}) as
\begin{multline}
\delta^2 (\lambda_p + \beta_1 (1+\beta_2\beta_4)\refspina - \beta_1 (\beta_2 + \beta_3)\refspinb) \\ = \frac{1}{\beta_1^2\beta_2 \la\refspina\refspinb\ra}\delta (\beta_3 - \frac{\la\refspina p\ra}{\beta_1\la\refspina\refspinb\ra} + \beta_2)\cdot
\delta (\beta_4 - \frac{\la\refspinb p\ra}{\beta_1\beta_2 \la\refspina\refspinb\ra} + \frac{1}{\beta_2})
\end{multline}
choosing $\refspina = \vll_p$, $\refspinb = \xi$ and taking integrations over $\beta_3$, $\beta_4$ we get
\begin{multline}
A^{*}_{k,n+1} = \frac{1}{\kappa^{*}}\la\xi p\ra \int
\frac{d^{k\times (n+2)} C}{\text{Vol}[GL(k)]}
\frac{d\beta_1 d\beta_2}{\beta_1\beta_2^2}\frac{\delta^{k\times 2} \left( C'
\cdot \vltuu \right)
\delta^{k\times 4} \left(C'\cdot \vleuu \right)
\delta^{(n+2-k)\times 2} \left(C'^{\perp}\cdot \vlluu \right)}{(1\cdots k)(2\cdots k+1)\cdots (n+2\cdots k-1)} ,
\end{multline}
where now
\begin{align}
C'_{n+1} &= -\beta_1 C_{n+1} + \beta_1\beta_2 C_{n+2},
&C'^{\perp}_{n+1} &= C^{\perp}_{n+1} + \frac{1+\beta_1}{\beta_1\beta_2} C^{\perp}_{n+2}, \nonumber \\
C'_{n+2} &= -\beta_1 C_{n+2} + \frac{1+\beta_1}{\beta_2} C_{n+1},
&C'^{\perp}_{n+2} &= C^{\perp}_{n+2} + \beta_2 C^{\perp}_{n+1},
\end{align}
and
\begin{align}\label{SpinorsInDeformadGrassmannian2}
&\vlluu_i = \vll_i, & i = 1,\ldots  n& , &\vlluu_{n+1} &= \lambda_p ,
&\vlluu_{n+2} &= \xi \nonumber \\
&\vltuu_i = \vlt_i, & i = 1,\ldots  n& ,
&\vltuu_{n+1} &= \frac{\la\xi |k}{\la\xi p\ra},
&\vltuu_{n+2} &= - \frac{\la p |k}{\la\xi p\ra} , \nonumber \\
&\vleuu_i = \vlet_i,  & i = 1,\ldots  n& , &\vleuu_{n+1} &= \vlet_p ,
&\vleuu_{n+2} &= 0 . \nonumber \\
\end{align}
Introducing inverse $C$-matrix transformation
\begin{align}
C_{n+1} &= C'_{n+1} +\beta_2 C'_{n+2} \nonumber \\
C_{n+2} &= \frac{1+\beta_1}{\beta_1\beta_2} C'_{n+1} + C'_{n+2}
\end{align}
minors of $C$-matrix containing both $n+1$ and $n+2$ columns when rewritten in terms of minors of $C'$-matrix acquire extra $-\frac{1}{\beta_1}$ factor. For example, for $(n+1\cdots k-2)$ minor we have
\begin{align}
(n+1\cdots k-2) = -\frac{1}{\beta_1} (n+1\cdots k-2)'\, .
\end{align}
This will generate total power of $(\beta_1)^{k-1}$ in the numerator.
Minors containing either $n+1$ or $n+2$ column transform as
\begin{align}
(n+2\, 1\cdots k-1) &= \frac{1+\beta_1}{\beta_1\beta_2}(n+1\, 1\cdots k-1)' + (n+2\, 1\cdots k-1)'\, , \\
(n-k+2\cdots n+1) &= (n-k+2\cdots n+1)' + \beta_2 (n-k+2\cdots n\, n+2)'\, ,
\end{align}
while all other minors remain unchanged $(\cdots) = (\cdots)'$. Finally, accounting for the Jacobian of transformation $\left(-\frac{1}{\beta_1}\right)^k$ we get
\begin{multline}
A^{*}_{k,n+1} = -\frac{\la\xi p\ra}{\kappa^{*}} \int
\frac{d^{k\times (n+2)} C'}{\text{Vol}[GL(k)]}
\frac{d\beta_1 d\beta_2}{\beta_1\beta_2}\delta^{k\times 2} \left( C'
	\cdot \vltuu \right)
	\delta^{k\times 4} \left(C'\cdot \vleuu \right)
	\delta^{(n+2-k)\times 2} \left(C'^{\perp}\cdot \vlluu \right) \nonumber \\
\times \frac{1}{(1\cdots k)'\cdots (n+2\cdots k-1)'\left(1+\beta_2\frac{(n-k+2\cdots n\, n+2)'}{(n-k+2\cdots n\, n+1)'}\right)\left(\beta_1\beta_2 + (1+\beta_1)\frac{(n+1\, 1\cdots k-1)}{(n+2\, 1\cdots k-1)}\right)}	
\end{multline}
Now, understanding integral over $\beta_{1,2}$ as residue form and taking $res_{\beta_1=-1}\circ res_{\beta_2=0}$
we recover our result from \cite{offshell-1leg}:
\begin{equation}
A^*_{k,n+1} = \int_{\Gamma_{k,n+2}^{tree}}\frac{d^{k\times
(n+2)}C'}{\text{Vol}[GL (k)]}Reg. \frac{\delta^{k\times 2} \left( C'
    \cdot \vltuu \right)
    \delta^{k\times 4} \left(C' \cdot \vleuu \right)
    \delta^{(n+2-k)\times 2} \left(C'^{\perp} \cdot \vlluu \right)}{(1 \cdots k)'\cdots (n+1 \cdots k-2)' (n+2 \; 1\cdots k-1)'}, \label{offshellAmpGrassmannianA}
\end{equation}
with
\begin{eqnarray}
    Reg.=\frac{\la\xi p\ra}{\kappa^{*}}\frac{(n+2 \; 1\cdots k-1)'}{(n+1 \; 1\cdots k-1)'}.
\end{eqnarray}

Now let's consider form factors of stress tensor supermultiplet operators. Let's consider (\ref{FormFactorsIntermidiate}).
In this case we can write it as
\begin{eqnarray}
F_{k,n} = \int\prod_{i=n+1}^{n+2}\frac{d^2\vll_i d^2\vlt_i}{\text{Vol}[GL(1)]} d^4\vlet_i F_{2,2}(\Omega_{n+1},\Omega_{n+2};\mathcal{T})\Big|_{\vll\to -\vll} A_{k,n+2} +\, \text{\it other gluing positions}.\, , \nonumber \\
\end{eqnarray}
where the minimal form factor $F_{2,2}(\Omega_{n+1},\Omega_{n+2};\mathcal{T})$ is given by (\ref{minimalSETformfactor}). Performing next on-shell integrations for particles $n+1$ and $n+2$ as above\footnote{See \cite{FormFactorsGrassmanians} for details.} we get (here we are considering only single term, corresponding to the gluing of minimal form factor between legs $1$ and $n$. Other terms come from gluing between legs $i$ and $i+1$, $i= 1\ldots n-1$)
\begin{multline}
F_{k,n} = -\la\xi_A\xi_B\ra^2\int\frac{\d\beta_1\d\beta_2}{(1-\beta_1\beta_2)^2}
\int\frac{d^{k\times (n+2)} C}{\text{Vol}[GL(k)]}
\frac{\delta^{k\times 2} \left( C'
	\cdot \vltuu \right)
	\delta^{k\times 4} \left(C'\cdot \vleuu \right)
	\delta^{(n+2-k)\times 2} \left(C'^{\perp}\cdot \vlluu \right)}{(1\cdots k)(2\cdots k+1)\cdots (n+2\cdots k-1)} ,
\end{multline}
where
\begin{align}
C'_{n+1} &= \frac{1}{1-\beta_1\beta_2} C_{n+1} + \frac{\beta_1}{1-\beta_1\beta_2} C_{n+2},
&C'^{\perp}_{n+1} &= C^{\perp}_{n+1} - \beta_2 C^{\perp}_{n+2}, \nonumber \\
C'_{n+2} &= \frac{1}{1-\beta_1\beta_2} C_{n+2} + \frac{\beta_2}{1-\beta_1\beta_2} C_{n+1},
&C'^{\perp}_{n+2} &= C^{\perp}_{n+2} - \beta_1 C^{\perp}_{n+1},
\end{align}
and
\begin{align}
&\vlluu_i = \vll_i, & i = 1,\ldots  n& , &\vlluu_{n+1} &= \xi_A ,
&\vlluu_{n+2} &= \xi_B \nonumber \\
&\vltuu_i = \vlt_i, & i = 1,\ldots  n& ,
&\vltuu_{n+1} &= -\frac{\la\xi_B |q}{\la\xi_B\xi_A\ra},
&\vltuu_{n+2} &= - \frac{\la\xi_A |q}{\la\xi_A\xi_B \ra} , \nonumber \\
&\vleuu_i^+ = \vlet_i^+,  & i = 1,\ldots  n& , &\vleuu^+_{n+1} &= 0 ,
&\vleuu^+_{n+2} &= 0 , \nonumber \\
&\vleuu^-_i = \vlet^-_i,  & i = 1,\ldots  n& , &\vleuu^-_{n+1} &= -\frac{\la\xi_B|\gamma^-}{\la\xi_B\xi_A\ra} ,
&\vleuu^-_{n+2} &=  -\frac{\la\xi_A|\gamma^-}{\la\xi_A\xi_B\ra}. \nonumber \\
\end{align}
The transition from the integration over $C$-matrix to integration over $C'$ matrix is again done similar to the case of off-shell amplitude considered above. This way our form factor is written as
\begin{multline}
F_{k,n} = -\la\xi_A\xi_B\ra^2  \int
\frac{d^{k\times (n+2)} C'}{\text{Vol}[GL(k)]}
\frac{d\beta_1 d\beta_2}{(1-\beta_1\beta_2)}\delta^{k\times 2} \left( C'
\cdot \vltuu \right)
\delta^{k\times 4} \left(C'\cdot \vleuu \right)
\delta^{(n+2-k)\times 2} \left(C'^{\perp}\cdot \vlluu \right) \nonumber \\
\times \frac{1}{(1\cdots k)'\cdots (n+2\cdots k-1)'\left(1-\beta_1\frac{(n-k+2\cdots n\, n+2)'}{(n-k+2\cdots n\, n+1)'}\right)\left(1 - \beta_2\frac{(n+1\, 1\cdots k-1)}{(n+2\, 1\cdots k-1)}\right)}	
\end{multline}
Finally, taking residues at $\beta_1 = \frac{(n-k+2\cdots n\, n+1)'}{(n-k+2\cdots n\, n+2)'}$ and $\beta_2 = \frac{(n+2\, 1\cdots k-1)'}{(n+1\, 1\cdots k-1)'}$ we reproduce the result of \cite{FormFactorsGrassmanians}:
\begin{equation}
F_{k,n} =  \int
\frac{d^{k\times (n+2)} C}{\text{Vol}[GL(k)]}
Reg.
\frac{\delta^{k\times 2} \left( C
	\cdot \vltuu \right)
	\delta^{k\times 4} \left(C\cdot \vleuu \right)
	\delta^{(n+2-k)\times 2} \left(C^{\perp}\cdot \vlluu \right)}{(1\cdots k)(2\cdots k+1)\cdots (n+2\cdots k-1)}\, ,
\end{equation}
where, now
\begin{equation}
Reg.=\la\xi_A\xi_B\ra^2\frac{Y}{1-Y},~Y = \frac{(n-k+2\cdots n\, n+1)(n+2\, 1\cdots k-1)}{(n-k+2\cdots n\, n+2)(n+1\, 1\cdots k-1)} .
\end{equation}
In the formula above we assumed a sum over different top-cell forms corresponding to different gluing positions of the minimal form factor. Note, that  string correlation function knows about these different top cells by construction.

\section{Gluing operator for the tree amplitudes}\label{appC}
In this appendix we present computational details for application of
gluing operator $\hat{A}_{i,i+1}$ to the tree level on-shell amplitudes.
First of all let's once more define $\hat{A}_{i,i+1}$:
\begin{eqnarray}
\hat{A}_{i,i+1}[f]\equiv\int\prod_{i=n+1}^{n+2} \frac{d^2\lambda_{i}d^2\tilde{\lambda}_{i}d^4\eta}{\mbox{Vol}[GL(2)]} ~A^*_{2,2+1}~\times~f\left(\{\lambda_i,\tilde{\lambda}_i,\eta_i\}_{i=1}^{n+2}\right),
\end{eqnarray}
Performing integration over  $\vlt_{n+1}$, $\vlt_{n+2}$, $\vlet_{n+1}$ and $\vlet_{n+2}$ variables as in Appendix \ref{appA} we get
\begin{eqnarray}
\hat{A}[f]=\frac{\langle p \xi\rangle}{\kappa^*}\int \frac{d\beta_1}{\beta_1}
\wedge
\frac{d\beta_2}{\beta_2}
~\frac{1}{\beta_1^2\beta_2} ~ f\left(\{\lambda_i,\tilde{\lambda}_i,\tilde{\eta}_i\}_{i=1}^{n+2}\right)\big{|}_{*},
\end{eqnarray}
where $\big{|}_{*}$ denotes substitutions $\{\lambda_i,\tilde{\lambda}_i,\eta_i\}_{i=n+1}^{n+2}
\mapsto\{\lambda_i(\beta),\tilde{\lambda}_i(\beta),\tilde{\eta}_i(\beta)\}_{i=n+1}^{n+2}$ with
\begin{align}
&\vll_{n+1}(\beta) = \vlluu_{n+1} + \beta_2\vlluu_{n+2}\, , && \vlt_{n+1}(\beta) =
\beta_1\vltuu_{n+1}  + \frac{(1+\beta_1)}{\beta_2}\vltuu_{n+2}\, ,
&&\vlet_{n+1}(\beta) = -\beta_1\vleuu_{n+1}\, , \nonumber\\
&\vll_{n+2}(\beta) = \vlluu_{n+2} + \frac{(1+\beta_1)}{\beta_1\beta_2}\vlluu_{n+1}\, ,
&& \vlt_{n+2}(\beta) = -\beta_1\vltuu_{n+2} -\beta_1\beta_2 \vltuu_{n+1}\, , &&\vlet_{n+2}(\beta) = \beta_1\beta_2\vleuu_{n+1}\, .
\end{align}
and
\begin{eqnarray}
\vlluu_{n+1}=\lambda_p,~\vltuu_{n+1}=\frac{\la\xi |k}{\la\xi p\ra},~\vleuu_{n}=\vlet_p;
~~\vlluu_{n+2}=\lambda_{\xi},~\vltuu_{n+2}=\frac{\la p |k}{\la\xi p\ra},~\vleuu_{n+2}=0.
\end{eqnarray}
We understand integration with respect to $\beta_{1,2}$ as residue form, and
will always evaluate it at points $res_{\beta_2=0}\circ
res_{\beta_1=-1}$.

The following formulas are useful in computations.
The transformed momenta for $n+1$'th and $n+2$'th particles are then given by
\begin{eqnarray}\label{momentaAsFunctiosFoBeta}
p_{n+1}(\beta)&=&
-\beta_1\vlluu_{n+1}\vltuu_{n+1}+\frac{1+\beta_1}{\beta_2}\vlluu_{n+1}\vltuu_{n+2}
-\beta_1\beta_2\vlluu_{n+2}\vltuu_{n+1}+(1+\beta_1)\vlluu_{n+2}\vltuu_{n+2},\nonumber\\
p_{n+2}(\beta)&=&
-\beta_1\vlluu_{n+2}\vltuu_{n+2}-\frac{1+\beta_1}{\beta_2}\vlluu_{n+1}\vltuu_{n+2}
+\beta_1\beta_2\vlluu_{n+2}\vltuu_{n+1}+(1+\beta_1)\vlluu_{n+1}\vltuu_{n+1}.\nonumber\\
\end{eqnarray}
and using definitions above  it is easy to see that
\begin{eqnarray}
k&=&\lambda_{n+1}(\beta)\tilde{\lambda}_{n+1}(\beta)
+\lambda_{n+2}(\beta)\tilde{\lambda}_{n+2}(\beta),\nonumber\\
\lambda_p\eta_p&=&\lambda_{n+1}(\beta)\tilde{\eta}_{n+1}(\beta)
+\lambda_{n+2}(\beta)\tilde{\eta}_{n+2}(\beta),
\end{eqnarray}
for all values of $\beta_1$ and $\beta_2$.

First of all let's make comment for the considered in the main text example (\ref{MHVGluingExample}) that the gluing operation commutes with projectors on particular physical particles provided
we identify $n+1$ and $n+2$  particles with gluons with $-+$ polarizations. Indeed from the previous example we have \cite{offshell-1leg,vanHamerenBCFW1} ($A_{2,4}\equiv A_{2,4}(\Omega_1,\ldots,\Omega_4)$)
\begin{eqnarray}
\partial_{\tilde{\eta}_2}^4\partial_{\tilde{\eta}_p}^4\hat{A}[A_{2,4}]=\hat{A}[\partial_{\tilde{\eta}_2}^4\partial_{\tilde{\eta}_3}^4A_{2,4}]=
\frac{\delta^4(p_{12}+k)}{\kappa^*\langle12\rangle}~res_{\beta_1=-1}\circ res_{\beta_2=0}[\omega]=A^*_{2,2+1}(1^+2^-g^*_3),\nonumber\\
\end{eqnarray}
where $\omega$ is given now by
\begin{eqnarray}
\omega=\frac{(\langle2p\rangle+\beta_2\langle2\xi\rangle)^3d\beta_2\wedge d\beta_1}{\beta_2\beta_1(\beta_1\beta_2\langle1\xi\rangle+(1+\beta_1)\langle1p\rangle)}.
\end{eqnarray}

Now let's proceed with more involved examples considered in the text and reproduce results for $A^*_{3,3+1}(1^-2^-3^+g_4^*)$ and $A^*_{3,4+1}(1^+2^+3^-4^-g_5^*)$ amplitudes from \cite{offshell-1leg}. In the case of $A^*_{3,3+1}(1^-2^-3^+g_4^*)$ amplitude we have to start with $A_{3,5}(1^-2^-3^+4^-5^+)$ amplitude (here and below $c^{-1}=\langle p \xi\rangle$):
\begin{eqnarray}
\partial_{\tilde{\eta}_1}^4\partial_{\tilde{\eta}_2}^4\partial_{\tilde{\eta}_4}^4A_{3,5}
(\Omega_1,\ldots,\Omega_5)=A_{3,5}(1^-2^-3^+4^-5^+)=\delta^4(p_{12345})
\frac{[35]^4}
{[12][23][34][45][51]},
\end{eqnarray}
so that
\begin{eqnarray}
A_{3,5}(1^-2^-3^+4^-5^+)\Big{|}_{*}=
\frac{\delta^4(p_{123}+k)~\beta_1^2\beta_2(\kappa^*c^{-1}[p3]+\beta_2[3\underline{\underline{4}}])^4}
{k^2c^{-1}[12][23](-\beta_1\beta_2[3\underline{\underline{4}}]+(1+\beta_1)c^{-1}\kappa^*[3p])
	([1p]c^{-1}\kappa^*+\beta_2[1\underline{\underline{4}}])}.\nonumber\\
\end{eqnarray}
Now, recalling that $k^2=-\kappa^*\kappa$ we get \cite{offshell-1leg}
\begin{eqnarray}
\hat{A}_{45}[A_{3,5}(1^-2^-3^+4^-5^+)
]=
\frac{\delta^4(p_{123}+k)[p3]^3}{\kappa [12][23][p1]}=A^*_{2,3+1}(1^-2^-3^+g_4^*),
\end{eqnarray}
and the integration with respect to $\beta$'s was performed by taking composite residue $res_{\beta_1=-1}\circ res_{\beta_2=0}[...]$.

In a similar fashion for $A_{3,5}(1^+2^+3^-4^-5^-6^+)$ amplitude we have
\begin{eqnarray}
\partial_{\tilde{\eta}_3}^4\partial_{\tilde{\eta}_4}^4\partial_{\tilde{\eta}_5}^4A_{3,5}
(\Omega_1,\ldots,\Omega_6)=
A_{3,5}(1^+2^+3^-4^-5^-6^+)=A+B,
\end{eqnarray}
with
\begin{eqnarray}
A&=&\frac{\la 3|1+2|6]^3}{[4 5][5 6]}\frac{\delta^4(p_{1\ldots6})}{\la 1 2\ra\la 2 3\ra p^2_{1,3} \la 1|2+3|4] },\nonumber\\
B&=&\frac{\la 5| 3+4|2]^3}{\la 5 6\ra\la 6 1\ra}\frac{\delta^4(p_{1\ldots6})}{ [2 3] [3 4] p_{2,4}^2 \la 1|2+3|4]}.
\end{eqnarray}
Next, it is not hard to see that ($[x|\equiv \la 3|(1+2)$, $|y\rangle \equiv (3+4)|2]$)
\begin{eqnarray}
A\Big{|}_{*}&=&\frac{\beta_1^2\beta_2([px]c^{-1}\kappa^*+\beta_2[\underline{\underline{5}}x])^3}
{c^{-1}\kappa\kappa^*(\beta_1^2\beta_2[4\underline{\underline{5}}]+(1+\beta_1)[4p]\kappa^*c^{-1})}
\frac{\delta^4(p_{1234}+k)}{\la 1 2\ra\la 2 3\ra p^2_{1,3} \la 1|2+3|4] },\nonumber\\
B\Big{|}_{*}&=&\frac{\beta_1^2\beta_2(\langle py\rangle +\beta_2\langle\xi y \rangle)^3}
{c~(\langle1\xi\rangle\beta_1\beta_2+(1+\beta_1)\langle p1\rangle)}\frac{\delta^4(p_{1234}+k)}{ [2 3] [3 4] p_{2,4}^2 \la 1|2+3|4]}.
\end{eqnarray}
Now, defining
\begin{eqnarray}
\omega_A&=&\frac{([px]c^{-1}\kappa^*+\beta_2[\underline{\underline{5}}x])^3}
{c^{-1}\kappa\kappa^*(\beta_1\beta_2[4\underline{\underline{5}}]+(1+\beta_1)[4p]\kappa^*c^{-1})}
\frac{d\beta_2\wedge d\beta_1}{\beta_1\beta_2},\nonumber\\
\omega_B&=&\frac{(\langle py\rangle +\beta_2\langle\xi y \rangle)^3}
{c~(\langle1\xi\rangle\beta_1\beta_2+(1+\beta_1)\langle p1\rangle)}\frac{d\beta_2\wedge d\beta_1}{\beta_1\beta_2},
,
\end{eqnarray}
we get
\begin{eqnarray}
\hat{A}_{56}[A]=\frac{\delta^4(p_{1234}+k)}{\la 1 2\ra\la 2 3\ra
	p^2_{1,3} \la 1|2+3|4] }
\frac{1}{c^{-1}\kappa^*}res_{\beta_1=-1}\circ res_{\beta_2=0}[\omega_A]=\frac{1}{\kappa}\frac{\delta^4(p_{1234}+k)\la 3|1+2|p]^3}{\la 1 2\ra\la 2 3\ra[4 p] p^2_{1,3} \la 1|2+3|4] },\nonumber\\
\end{eqnarray}
and
\begin{eqnarray}
\hat{A}_{56}[B]=\frac{\delta^4(p_{1234}+k)}{ [2 3] [3 4] p_{2,4}^2 \la
	1|2+3|4]} \frac{1}{c^{-1}\kappa^*}res_{\beta_1=-1}\circ
res_{\beta_2=0}[\omega_B]=\frac{1}{\kappa^{*}}
\frac{\delta^4(p_{1234}+k)\la p| 3+4|2]^3}{\la p 1\ra [2 3] [3 4] p_{2,4}^2 \la 1|2+3|4]},\nonumber\\
\end{eqnarray}
So, as expected \cite{offshell-1leg}
\begin{eqnarray}
\hat{A}_{56}[A_{3,5}(1^+2^+3^-4^-5^-6^+)]=A^{*}_{3,4+1}(1^+2^+3^-4^-g_5^*).
\end{eqnarray}

As a final example let's consider the following example. Let's reproduce $A^*_{3,0+3}(g_1^*,g_2^*,g_3^*)$ Wilson line correlation function from $A_{3,6}(1^-2^+3^-4^+5^-6^+)$ on-shell amplitude.
According to our previous discussion $A^*_{3,0+3}$ could be written as
\begin{equation}
A^*_{3,0+3}(g_1^*,g_2^*,g_3^*)
=(\hat{A}_{12}\circ\hat{A}_{34}\circ\hat{A}_{56})[A_{3,6}(1^-2^+3^-4^+5^-6^+)]], ,
\end{equation}
where $A_{3,6}(1^-2^+3^-4^+5^-6^+)$ amplitude is given by
\begin{equation}
A_{3,6}=\delta^4(p_{1\ldots6})
\left(1+\mathbb{P}^2+\mathbb{P}^4\right)f,
~f=\frac{\langle13\rangle^4[46]^4}{\langle12\rangle\langle23\rangle[45][56]\langle3|1+2|6]\langle1|5+6|4]p_{456}^2}
\end{equation}
and $\mathbb{P}$ is permutation operator shifting spinor labels by +1 mod 6. The algebraic manipulation related to the actions of $\hat{A}_{ii+1}$ operators are identical to those already discussed. The factors $1/\beta^2_1\beta_2$ in the definition of gluing operators will cancel with corresponding factors in the amplitude after substitutions applied, while integrals are evaluated by composite residues  $res_{\beta_1=-1}\circ res_{\beta_2=0}$. So, in what follows we will present only the results of applying gluing operators $\hat{A}_{ii+1}$ to on-shell amplitude. For $f$ term we have:
\begin{equation}
\hat{A}_{56}[f]=\delta^4(p_{1234}+k_3)
\frac{\langle13\rangle^4[4p_3]^3}{\kappa_3\langle12\rangle\langle23\rangle\langle3|1+2|p_3]\langle1|k_3|4]p_{123}^2},
\end{equation}
and
\begin{equation}
(\hat{A}_{34}\circ\hat{A}_{56})[f]=\delta^4(p_{12}+k_2+k_3)
\frac{\langle1p_2\rangle^4[p_2p_3]^3}{\kappa_3\langle12\rangle\langle2p_2\rangle\langle
	p_2|1+2|p_3]\langle1|k_3|p_2]\langle p_2|k_3|p_2]}.
\end{equation}
Note that the ordinary propagator $1/p_{123}^2$ transformed into eikonal one $1/(p_2k_3)$ after the action of gluing operator. Finally
\begin{equation}
(\hat{A}_{12}\circ\hat{A}_{34}\circ\hat{A}_{56})[f]=\delta^4(k_1+k_2+k_3)
\frac{\langle p_1p_2\rangle^3[p_2p_3]^3}{\kappa_3\kappa^*_1\langle
	p_2|k_1|p_3]\langle p_1|k_3|p_2]\langle p_2|k_1|p_2]},
\end{equation}
where we used that $\langle p_2|k_3|p_2]=\langle p_2|k_1|p_2]$. Other
terms can be obtained by similar computations or just by careful relabeling of indexes. The final result takes the form:
\begin{eqnarray}
A^*_{3,0+3}&=&(\hat{A}_{12}\circ\hat{A}_{34}\circ\hat{A}_{56})[A_{3,6}(1^-2^+3^-4^+5^-6^+)]=\delta^4(k_1+k_2+k_3)
\left(1+\mathbb{P}'+\mathbb{P}'^2\right)\tilde{f},\nonumber\\
\tilde{f}&=&\frac{\langle
	p_1p_2\rangle^3[p_2p_3]^3}{\kappa_3\kappa^*_1\langle p_2|k_1|p_3]\langle
	p_1|k_3|p_2]\langle p_2|k_1|p_2]}.
\end{eqnarray}
Here $\mathbb{P}'$ is permutation operator which now  shifts all spinor and momenta labels by +1 mod 3. Obtained expression is in full agreement with previous computations using both Grassmannian integral representation \cite{offshell-multiplelegs} and BCFW recursion \cite{vanHamerenBCFW1}.

\bibliographystyle{ieeetr}
\bibliography{refs}

\end{document}